\documentclass[aps,prb,,twocolumn,showpacs,superscriptaddress]{revtex4}

\usepackage{amsmath,graphicx,amsbsy,amssymb,amsmath,epsfig,latexsym,wasysym,pifont,mathrsfs,natbib}
\usepackage{float}
\newfloat{widefig}{thp}{lop}
\usepackage{graphicx}
\usepackage{comment}
\usepackage{verbatim}
\usepackage{multirow,array}
\usepackage{hyperref}
\usepackage{subfigure}
\usepackage{color}

\begin{document}

\title{Systematic understanding of half-metallicity of ternary compounds in Heusler and Inverse Heusler structures with 3$d$ and 4$d$ elements. }

\author{Srikrishna Ghosh}
\affiliation{Department of Physics, Indian Institute of Technology Guwahati, Guwahati-781039, Assam, India.} 
\author{Subhradip Ghosh}
\email{subhra@iitg.ac.in}
\affiliation{Department of Physics, Indian Institute of Technology Guwahati, Guwahati-781039, Assam, India.}

\date{\today}
\begin{abstract}
Employing {\it ab initio} electronic structure calculations we extensively study ternary  Heusler compounds having the chemical formula X$_2$X$^\prime$Z, where X = Mn, Fe or Co; Z = Al or Si; and X$^\prime$ changes along the row of 4$d$ transition metals. A comprehensive overview of these compounds, addressing the trends in structural, electronic, magnetic properties and Curie temperature is presented here along with the search for new materials for spintronics applications. A simple picture of hybridization of the $d$ orbitals of the neighboring atoms is used to explain the origin of the half-metallic gap in these compounds. We show that arrangements of the magnetic atoms in different Heusler lattices are largely responsible for the interatomic exchange interactions  that are correlated with the features in their electronic structures as well as possibility of half-metallicity. We find seven half-metallic magnets with 100\% spin polarization. We identify few other compounds with high spin polarisation as ``near half-metals" which could be of potential use in applications as well. We find that the major features in the electronic structures remain intact if a 3$d$ X$^{\prime}$ constituent is replaced with an isoelectronic 4$d$, implying that the total number of valence electrons can be used as a predictor of half-metallic nature in compounds from Heusler family.

\end{abstract}

\maketitle
\section{INTRODUCTION}
Since the discovery in 1903, Heusler alloys have been exhaustively studied for its numerous potential applications. A variety of diverse magnetic phenomena ranging from half-metallicity, magnetic shape-memory effect, giant magnetocaloric effect, thermoelectric effect, spin-light emitting diode \cite{katsnelson2008mi,PhysRevB.66.134428,PonsMSE08,GalanakisPRB02,AntoniJPCM09,DoPRB11,Co2FeSi_3} to recently discovered Spin gapless semiconductor, topological insulators, superconductivity, multifunctional  Heuslers for recording \cite{wang2008sgs,wang2010zero,Felser_superconductivity,ouardi2011topological,kurt2011mn3ga}  have been identified in some of the Heusler groups, comprising more than thousand compounds. 

Among the explored intermetallics, full Heuslers (L2$_1$) X$_2$YZ are largest in number, where X and Y are transition metal and Z is  a main group element. It is obvious that the unfilled $d$-shells of more than one transition metal elements give rise to the novel magnetic properties of these Heusler alloys. The occupancies of the transition metal magnetic elements in  sub-lattices with various symmetries, the magnetic structure associated with and the resulting electronic structures are responsible for the novel properties in these materials.

Search for half-metals (HM) and spin gapless semiconductors (SGS) has been one of the forerunning research area in condensed matter physics due to these materials being suitable for spintronics applications. Extensive investigations on Heusler compounds, towards this direction, has paid rich dividends as quite a few HM and SGS have been discovered with composition X$_{2}$YZ where X and Y are elements from the 3$d$ series of periodic table \cite{AlijaniJAP13, MeshcheriakovaPRL14,VajihehJPCM12,GalanakisJAP14, EndoJAC12, GrafPSSC11, WollmannPRB14}. The investigations into compounds with either both X and Y are elements with 4$d$ electrons or one of them a 4$d$ element are fewer in number. Presence of both 3$d$ and 4$d$ magnetic elements in the same compound with Heusler structure is expected to provide interesting perspectives. However, there is no systematic investigations available in Heusler compounds having both 3$d$ and 4$d$ elements.

In this paper, we, therefore, have tried to explore six ternary series X$_2$X$^\prime$Z, with X = Mn, Fe, Co , Z = Al, Si while X$^\prime$ is varied from Y to Ag down the series in the periodic table by Density functional theory (DFT) based electronic structure calculations. Among the family of ternary Heusler compounds Co$_2$ and Mn$_2$-based  Heusler compounds have drawn commendable interest for applications in the areas of spintronics and other magnetism related applications due to their high Curie temperatures \cite{wurmehl2006investigation, GrafPSSC11, weht1999half, csacsioglu2005first}. Fe-based Heusler compounds are known to be  soft ferromagnets with Curie temperature as high as 900 K making them suitable for potential magnetic applications \cite{gasi2013iron}. Thus, we find it intriguing to explore specifically Mn$_2$ , Fe$_2$ and Co$_2$ based ternary alloys for the search of new materials with one magnetic element being one with 4$d$ electrons, for spintronics applications. Apart from the quest for new half-metallic magnets our motivations for this work are the following:

(1) to understand the effects and their origin, of the presence of an element with $4d$ electrons on the structure, electronic and magnetic properties of ternary  compounds where the other magnetic elements are from $3d$ transition metal series. A systematic study spanning across several Heusler series would help understand the comparative roles of 4$d$ and 3$d$ elements.

(2) to understand the effects and their origin, of replacing an element with $3d$ electrons by an element with $4d$ electrons on the electronic and magnetic properties in the context of half-metallicity.

(3) to understand the impacts and their origin, of replacing the main group element  Si with  Al in case of ternary alloys having both 3$d$ and 4$d$ elements as magnetic components.

The paper is organised as follows:  Computational details and the methods, used in this work are given in the next section. In the subsequent  section (Section III) we present the discussions of the calculated results on structural, electronic and magnetic properties of the ternary compounds mentioned above along with a comparative understanding with compounds having only 3$d$ elements as the magnetic components. The conclusions and future outlook is presented at the end. 



\section{Computational Details} 

We have used spin-polarized DFT based projector augmented wave methods as implemented in Vienna Ab-initio Simulation Package (VASP) \cite{PAW94,VASP196,VASP299} with Generalized Gradient Approximation (GGA) \cite{PBEGGA96} for calculating electronic structure. An energy cut-off of 450 eV and a Monkhorst-Pack \cite{MP89} 25 x 25 x 25 k-mesh was used for self-consistent calculations while a larger 31 x 31 x 31 k-mesh was used for calculating densities of sates. We set the energy and the force convergence criteria to $10^{-6}$ eV and $10^{-2}$ eV/\AA respectively. 

With multiple scattering Green function formalism as implemented in SPRKKR code\cite{EbertRPP11} we have calculated the magnetic pair exchange parameters. Here, the spin part of the Hamiltonian is mapped to a Heisenberg model

\begin{eqnarray*}
H = -\sum_{\mu,\nu}\sum_{i,j}
J^{ij}_{\mu\nu}
\mathbf{e}^{i}_{\mu}
.\mathbf{e}^{j}_{\nu}
\end{eqnarray*}
where $\mu$, $\nu$ represent different sub-lattices, \emph{i}, \emph{j} represent atomic positions and $\mathbf{e}^{i}_{\mu}$ denotes the unit vector along the direction of magnetic moments at site \emph{i} belonging to sub-lattice $\mu$. The $J^{ij}_{\mu\nu}$ are calculated from the energy differences due to infinitesimally small orientations of a pair of spins within the formulation of Liechtenstein {\it et. al.}\cite{LiechtensteinJMMM87}. Full potential spin polarized scaler relativistic Hamiltonian with angular momentum cut-off $l_{max} = 3$ is used to calculate the energy differences by the SPRKKR code. We have use a uniform k-mesh of $22\times22\times22$ for Brillouin zone integrations. The Green's functions were calculated for 32 complex energy points distributed on a semiconductor contour. For the self-consistence cycles, the energy convergence criterion was set to 10$^{-6}$ Ry . The  calculated $J^{ij}_{\mu\nu}$ were used further to compute the Curie temperatures within the mean field approximation\cite{SokolovskiyPRB12}.

\section{Results and Discussions}

\subsection{Structural Properties}

\begin{figure}[h!]
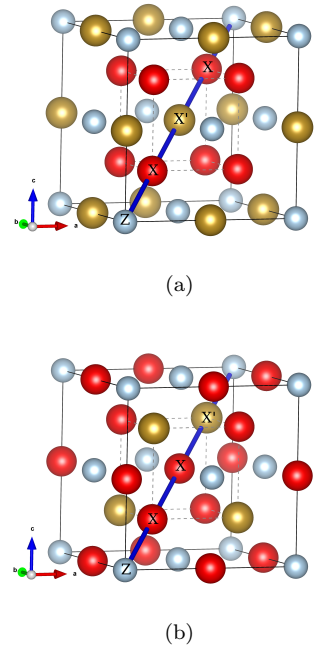

\centering

\subfigure[]{\includegraphics[scale=0.037]{fig1a.pdf}}
\subfigure[]{\includegraphics[scale=0.037]{fig1b.pdf}}

\caption{Crystal structure of ternary Heusler compounds X$_2$X$^\prime$Z in ({\bf a}) T$_{\text{I}}$ (Heusler {\it i.e} two X atoms are in symmetric positions) ({\bf b}) T$_{\text{II}}$ (Inverse Heusler {\it i.e} X and $\text{X}^\prime$ are in symmetric positions). }
\label{Crystal1}
\end{figure}

 X$_2$X$^{\prime}$Z ternary Heusler alloys crystallise in two prototype structures. The``regular'' Heusler or L2$_1$ (Cu$_2$MnAl prototype) crystallises in a cubic structure with the space group $Fm\bar{3}m$ (space group number 225). The other one, the  "inverse" Heusler, has the prototype Hg$_2$CuTi (space group number 216; $F\bar{4}3m$) structure. The structures  are schematically shown in Fig \ref{Crystal1}.

In L2$_1$ structure the Wyckoff positions $4a(0,0,0)$ and $4b(1/2, 1/2, 1/2)$ are occupied by the least and the most electronegative elements X$^{\prime}$ and Z respectively forming a rocksalt (NaCl) type lattice. The $8c$ Wyckoff positions (1/4, 1/4, 1/4) are occupied by two X atoms as shown in Fig.\ref{Crystal1}.(a). Since the interaction between X$^\prime$ and Z atoms are ionic in character, they are octahedrally coordinated. The tetrahedral voids are filled up with X atoms, having electronegativity in between those of X$^\prime$ and Z. In inverse Heusler structure, X atoms of X$_2$X$^\prime$Z compounds sit at the Wyckoff positions $4b(1/2, 1/2, 1/2)$ and $4c(1/4, 1/4, 1/4)$, whereas $4a(0,0,0)$ and $4d(3/4, 3/4, 3/4)$ are occupied by Z and X$^\prime$ atoms respectively \cite{WollmannPRB14} as shown in Fig.\ref{Crystal1}.(b).

In Table ~\ref{table2} we  present the structural properties (the crystal structure type and the lattice constant) of the lowest energy Heusler phases, obtained by our calculations, for the compounds considered here. The trends in the structure types can be understood from the following empirical rule based on the electronegativities of the constituents which has been successful in explaining the structural preferences for a number of Heusler intermetallics\cite{AlijaniPRB11,BainslaPRB215,EnamullahPRB15,KlaerPRB11,DaiJAP09,CoFeMnSi_aftab,BerriJMMM14,GaoSM15, XiongJMMM14,IzadiJSNM11,MukhtiyarJAC13,BainslaJMMM15,GokhanSSS12}: if the main group element is fixed at $4a$ site, the $4b$ site should be occupied by the least electronegative element of the remaining three transition metal atoms \cite{NSRP}.  From our results, we find that except Mn$_2$NbZ, X$_2$Mo$^\prime$Z (X = Mn, Fe, Co) and Co$_2$TcAl this rule explains the structure types for all other compounds. The computed lattice constants agree well with the existing results. In case of lattice constants, we find that the lattice constants of ternary Heuslers are greater when X$^{\prime}$ is an element with 4$d$ electrons, in comparison to it's counterpart with X$^{\prime}$ an element with 3$d$ electrons located right above the column in periodic table (Table I, supplementary material). This trend can be explained purely on the basis of relative sizes of the atomic radii of the X$^{\prime}$ component.

In order to first verify whether a compound indeed can be formed, we have calculated the formation energies the following way:
 
 \[E_f = E_{X_2X^{\prime}Z} - (2E_X + E_{X^\prime} + E_Z) \]
 
 Where $E_{X_2X^{\prime}Z}$ is the total energies per formula unit of the X$_2$X$^{\prime}$Z and $E_X$, $E_{X{^\prime}}$, $E_Z$, are the total energies of the bulk X, X$^\prime$ and Z respectively in their ground state structures. The results are tabulated in Table \ref{table2}. Except Fe$_2$AgAl, Mn$_2$AgSi and Fe$_2$AgSi most compounds are likely to form as the formation energies are negative and hence are thermally stable from the point of view of Enthalpy. Thus, it is worth investigating further the properties of all these ternary Heusler compounds. 

\begin{table*}
\caption{\label{table2}Calculated lattice constants, formation energies, structure type and magnetic moments of ternary X$_2$X$^{\prime}$Z compounds.} 
\begin{tabular}
{ l@{\hspace{0.5cm}} l@{\hspace{0.5cm}}  c@{\hspace{0.5cm}} c@{\hspace{0.5cm}} c@{\hspace{0.5cm}} l }

\hline\hline
Systems & Lattice  	& Formation   	    & Structure & M   \\
(Type-I)& constant(\AA) & energy(eV) 	    &  type     & ($\mu_{B}/f.u.$)  \\ [0.5ex]
\hline
Mn$_2$YAl & 6.52 & -0.74 & T$_\text{I}$ & 5.72\\
Mn$_2$ZrAl & 6.15 & -2.01 & T$_\text{I}$ & 3.00\\
Mn$_2$NbAl & 6.00 (6.005\cite{Mn2NbAl_kervan2016}) & -1.56 & T$_\text{I}$ & 2.00 (2.00\cite{Mn2NbAl_kervan2016})\\
Mn$_2$MoAl & 5.90 & -1.23 & T$_\text{I}$ & 1.04\\
Mn$_2$TcAl & 5.95 & -1.94 & T$_\text{II}$ & 0.04\\
Mn$_2$RuAl & 5.95 & -2.62 & T$_\text{II}$ & 1.01\\
Mn$_2$RhAl & 5.95 & -2.08 & T$_\text{II}$ & 1.86\\
Mn$_2$PdAl & 6.05 & -1.67 & T$_\text{II}$ & 0.86\\
Mn$_2$AgAl & 6.20 & -0.29 & T$_\text{II}$ & 0.31\\
\hline
Fe$_2$YAl & 6.25 & -0.47 & T$_\text{I}$ & 1.95\\
Fe$_2$ZrAl & 6.05 & -1.96 & T$_\text{I}$ & 0.91\\
Fe$_2$NbAl & 5.90 (5.909\cite{Hamad2016})  & -1.63 & T$_\text{I}$ & 0.00\\
Fe$_2$MoAl & 5.85 & -0.77 & T$_\text{I}$ & 0.82\\
Fe$_2$TcAl & 5.85 & -1.34 & T$_\text{II}$ & 3.29\\
Fe$_2$RuAl & 5.90 & -1.97 & T$_\text{II}$ & 5.20\\
Fe$_2$RhAl & 5.90 & -1.58 & T$_\text{II}$ & 5.00\\
Fe$_2$PdAl & 5.95 & -1.25 & T$_\text{II}$ & 4.84\\
Fe$_2$AgAl & 6.05 & 0.25 & T$_\text{II}$ & 4.75\\
\hline
Co$_2$YAl & 6.20 & -3.21 & T$_\text{I}$ & 0.00\\
Co$_2$ZrAl & 6.05 & -4.32 & T$_\text{I}$ & 1.00\\
Co$_2$NbAl & 5.95 & -3.31 & T$_\text{I}$ & 2.00\\
Co$_2$MoAl & 5.90 & -2.36 & T$_\text{I}$ & 2.84\\
Co$_2$TcAl & 5.85 & -3.14 & T$_\text{I}$ & 3.91\\
Co$_2$RuAl & 5.85 & -3.38 & T$_\text{II}$ & 3.87\\
Co$_2$RhAl & 5.85 & -2.84 & T$_\text{II}$ & 3.37\\
Co$_2$PdAl & 5.90 & -2.56 & T$_\text{II}$ & 3.04\\
Co$_2$AgAl & 5.97 & -1.28 & T$_\text{II}$ & 2.77\\
\hline
\hline
Mn$_2$YSi & 6.16 & -1.01 & T$_\text{I}$ & 3.02\\
Mn$_2$ZrSi & 6.00 (6.004 \cite{Mn2ZrSi_abada2015})& -2.42 & T$_\text{I}$ & 2.00\cite{Mn2ZrSi_abada2015}\\
Mn$_2$NbSi & 5.87 & -2.02 & T$_\text{I}$ & 0.99\\
Mn$_2$MoSi & 5.78 & -1.61 & T$_\text{I}$ & 0.00\\
Mn$_2$TcSi & 5.83 & -2.60 & T$_\text{II}$ & 1.04\\
Mn$_2$RuSi & 5.79 & -3.10 & T$_\text{II}$ & 2.00\\
Mn$_2$RhSi & 5.81 (5.905 \cite{PKJhaPhysica18})& -2.09 & T$_\text{II}$ & 3.00 (3.00 \cite{PKJhaPhysica18})\\
Mn$_2$PdSi & 5.98 & -1.39 & T$_\text{II}$ & 0.95\\
Mn$_2$AgSi & 6.11 & 0.14 & T$_\text{II}$ & 0.59\\
\hline
Fe$_2$YSi & 6.08 & -0.87 & T$_\text{I}$ & 0.96\\
Fe$_2$ZrSi & 5.93 (5.899 \cite{Fe2ZrSi}) & 2.55 & T$_\text{I}$ & 0.00\\
Fe$_2$NbSi & 5.84 & -1.58 & T$_\text{I}$ & 0.97\\
Fe$_2$MoSi & 5.78 & -0.71 & T$_\text{I}$ & 1.79\\
Fe$_2$TcSi & 5.79 & -1.88 & T$_\text{II}$ & 3.78\\
Fe$_2$RuSi & 5.78 & -2.37 & T$_\text{II}$ & 4.69\\
Fe$_2$RhSi & 5.80 & -1.59 & T$_\text{II}$ & 4.90\\
Fe$_2$PdSi & 5.88 & -0.91 & T$_\text{II}$ & 5.07\\
Fe$_2$AgSi & 5.97 & 0.77 & T$_\text{II}$ & 4.69\\
\hline
Co$_2$YSi & 6.12 & -3.32 & T$_\text{I}$ & 1.00\\
Co$_2$ZrSi & 5.99 & -4.36 & T$_\text{I}$ & 2.00\\
Co$_2$NbSi & 5.87 & -3.01 & T$_\text{I}$ & 1.71\\
Co$_2$MoSi & 5.78 & -2.16 & T$_\text{II}$ & 0.01\\
Co$_2$TcSi & 5.72 & -3.22 & T$_\text{II}$ & 0.21\\
Co$_2$RuSi & 5.74 & -3.60 & T$_\text{II}$ & 3.20\\
Co$_2$RhSi & 5.76 & -2.65 & T$_\text{II}$ & 3.27\\
Co$_2$PdSi & 5.79 & -2.14 & T$_\text{I}$ & 2.35\\
Co$_2$AgSi & 5.83 & -1.20 & T$_\text{I}$ & 0.20\\ \\ [1ex]
\hline\hline
\end{tabular}
\end{table*}


\subsection{The Magnetic moments and the Slater-Pauling Rule}

\begin{figure}[h!]
\centering
\subfigure[]{\includegraphics[scale=0.25]{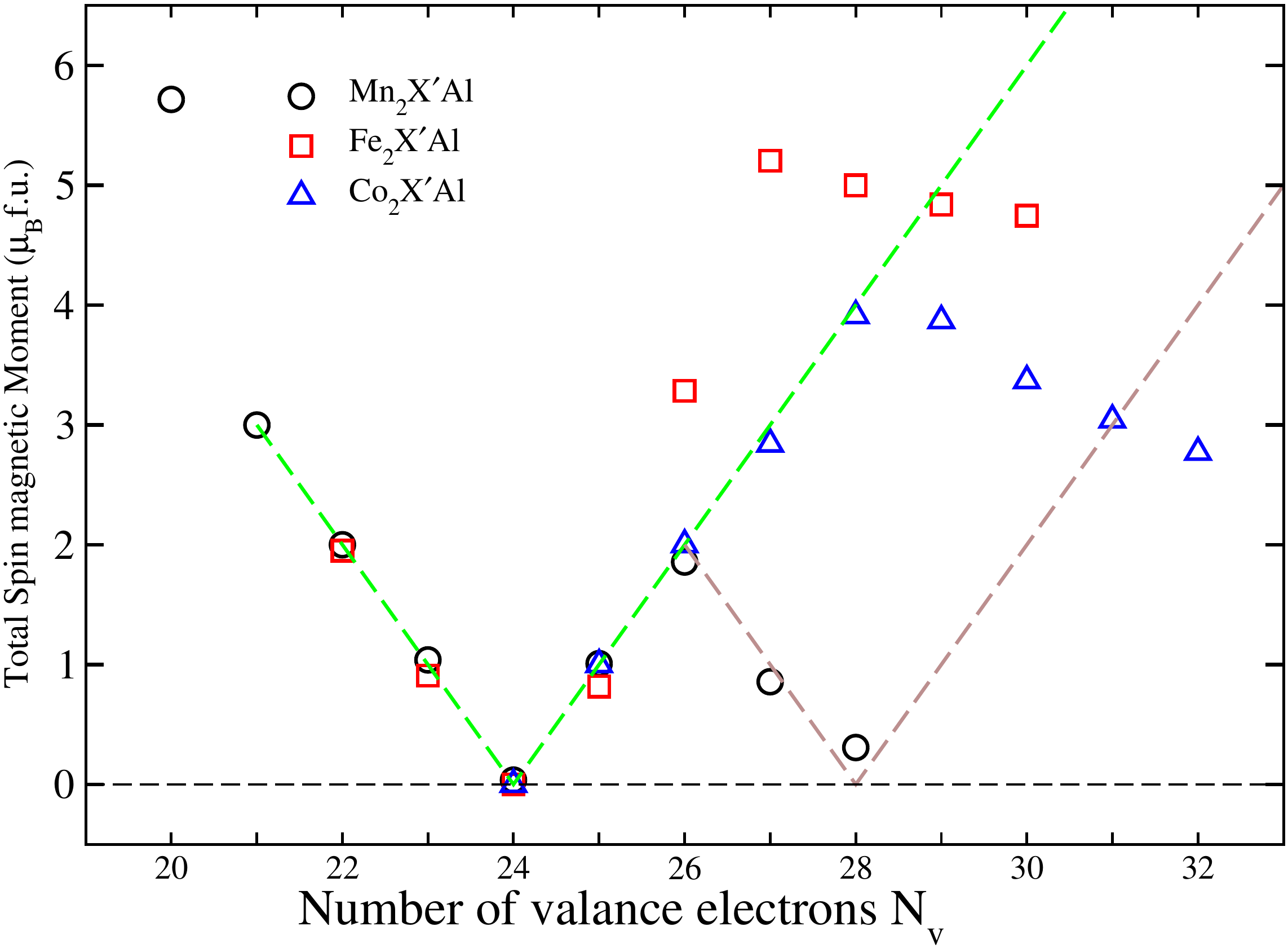}}
\subfigure[]{\includegraphics[scale=0.25]{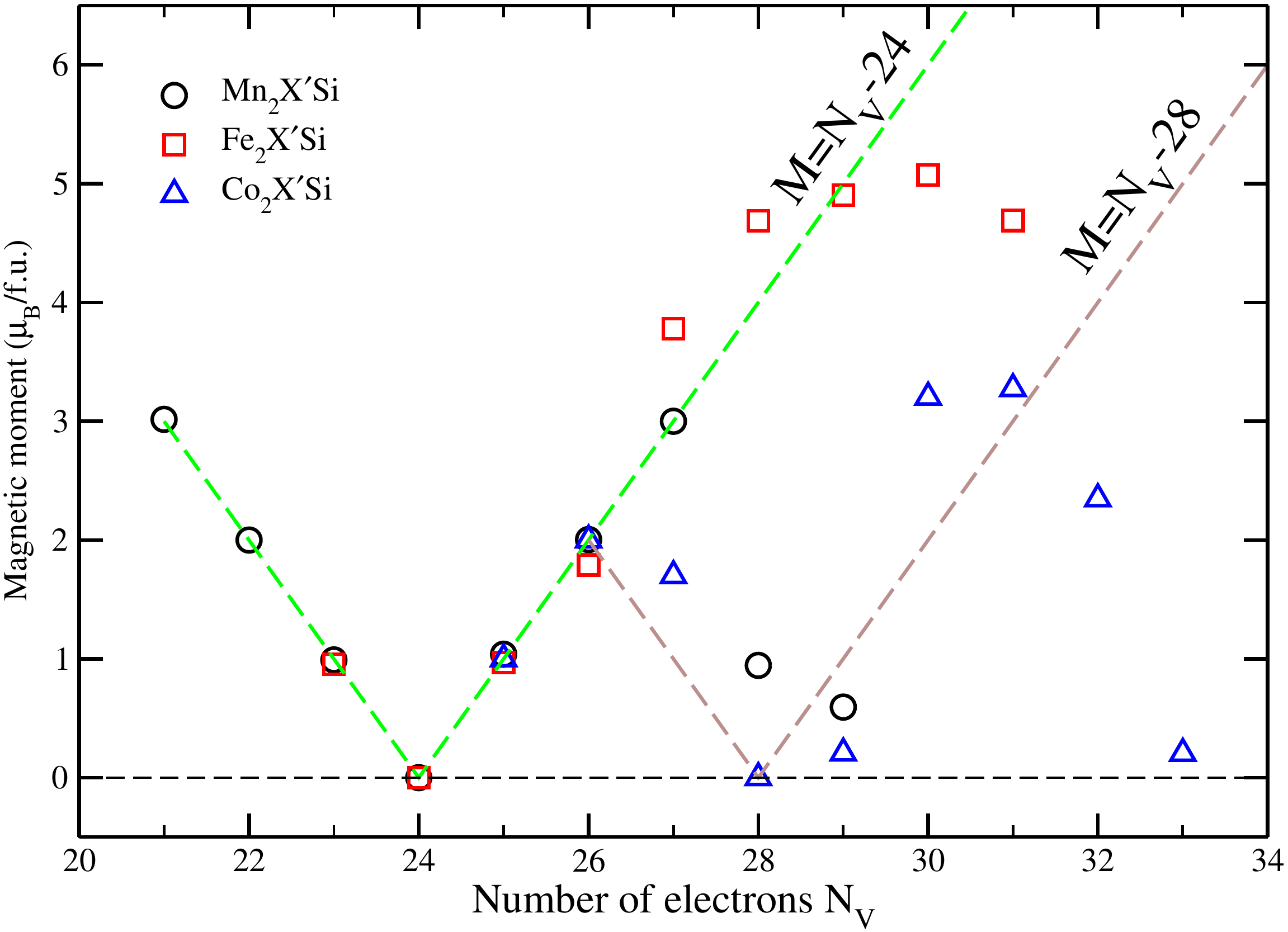}}
\caption{Total magnetic moments versus the total number of 
             valence electrons $N_{\text{V}}$ for  ({\bf a}) X$_2$X$^\prime$Al  and ({\bf b}) X$_2$X$^\prime$Si compounds respectively. The Slater-Pauling
             $M=N_{\text{V}}-24$ and $M=N_{\text{V}}-28$ lines are drawn as a guide to understand whether the compounds follow
             Slater-Pauling rule. }
\label{sp-rule}
\end{figure}

The necessary condition for a Heusler alloy to be half-metal is that the total spin-magnetic moment per formula-unit is an integer and that it follows the Slater-Pauling rule, connecting the magnetic moment to the total number of valance electrons per formula unit \cite{SlaterPRB36,PaulingPRB38,Galanakis_PRB13,Galanakis2016}. In case of half-metallic ternary Heusler alloys X$_{2}$X$^{\prime}$Z, the manifestation of this rule is that the total magnetic moment per formula unit, M and the number of valance electrons per formula unit, N$_{\text{V}}$ are related by M =$|\text{N}_{\text{V}} -18|$ or by M =$|\text{N}_{\text{V}} -24|$ or by M =$|\text{N}_{\text{V}} -28|$, depending on whether X is an early transition metal; or more precisely  whether the Fermi-energy is placed in a gap after 9, 12 or 14 electronic states respectively \cite{GalanakisPRB02,Galanakis_PRB13}.In Fig \ref{sp-rule}.(a), we have plotted the variations of M with N$_{\text{V}}$ for X$_2$X$^\prime$Al (X$^\prime = 4d$, X = Mn, Fe, Co).  Our results show that the moments of Mn$_2$X$^\prime$Z compounds follow the M =$|\text{N}_{\text{V}} -24|$ rule with the exceptions of Mn$_2$YAl, Mn$_2$PdZ and Mn$_2$AgZ where the expected moments are smaller than those predicted by the Slater-Pauling rule. However, the moment of Mn$_2$PdAl lies close to M =$|\text{N}_{\text{V}} -28|$ line. For Fe$_2$X$^\prime$Z series we find that for compounds with  X$^\prime$ an element after Mo, the moments of all compounds are much deviated from any of the Slater-Pauling lines. In case of Co$_2$X$^\prime$Al series the moments deviate from the line M =$|\text{N}_{\text{V}} -24|$ for late transition metals (X$^\prime$ = Ru, Rh, Pd, Ag), though Co$_2$PdAl lies close to M =$|\text{N}_{\text{V}} -28|$ line. For the case of Co$_2$X$^\prime$Si series except Co$_2$YSi and Co$_2$ZrSi none of the compounds follow the M =$|\text{N}_{\text{V}} -24|$ rule. Co$_2$MoSi with 28 valance electrons and having total moment zero lies on M =$|\text{N}_{\text{V}} -28|$. Across the six series under investigation there are only two compounds, Mn$_{2}$ZrAl and Mn$_{2}$YSi having moments in conformation with both M =$|\text{N}_{\text{V}} -18|$ and M =$|\text{N}_{\text{V}} -24|$ rules. In principle these system can have gaps after both 9 states and 12 states \cite{Dipanjan}.

Fig \ref{sp-rule} and Table \ref{table2} demonstrate that there are quite a few compounds with integer or close to integer moments following Slater-Pauling rule. These are the potential half-metals whose electronic structures need to be examined carefully before coming to any conclusion. In the next sub-section we discuss in detail the electronic structures of compounds in each series and try to understand the origin of half-metallicity in compounds with 4$d$ elements.

\subsection{Electronic Structure}

The origin of the half-metallic gap and the conformation to the Slater-Pauling rule by magnetic Heusler and inverse Heusler alloys can be understood from their electronic structures near the Fermi level by addressing the hybridisations between the constituents of the compounds \cite{GalanakisPRB02,Galanakis_PRB13,GalanakisJAP13_113}. In a ternary Heusler compound X$_{2}$X$^{\prime}$Z, the $sp$-element Z  provides 4 low lying (below $d$ bands) states-one $s$ and three $p$- in each spin channel. These $s-p$ bands lie well below the Fermi-energy, accommodate charges from the atoms with $d$ electrons and stabilize the structure\cite{PhysRevB.66.134428}. The $d$-orbitals of transition-metal atoms hybridise the following way: first the atoms on symmetric sites ($8c$ in the space group 225, $4c$ and $4d$ in space group-216) hybridise creating $(i)$ 5 bonding $d$-hybrids (2 doubly-degenerate $e_g$ and 3 triply degenerate $t_{2g}$ hybrid orbitals) $(ii)$ five non bonding hybrid $d$-orbitals (2 doubly-degenerate $e_u$ and 3 triply degenerate $t_{1u}$ hybrid orbitals)(Fig. 1(a) in Supplementary material). The bonding $e_g$ and $t_{2g}$ orbitals, having tetrahedral symmetry, can only hybridise with $d$-orbitals of the remaining transition metal atom. Thus, ultimately we are left with five bonding and five anti-bonding states which are non-transforming with the $u$ representation\cite{GalanakisPRB02,bouckaert1936lp} and can not couple with $d$-orbitals of the remaining transition atom (Fig. 1(b) in Supplementary material).

 In case of a half-metal there is an energy gap in one spin-channel, where the Fermi-energy is pinned in. The position of the Fermi-level in a spin-channel where the half-metallic gap is situated, determines which Slater-Pauling rule the material follows as the number of states in that particular spin-channel below the Fermi-level is always fixed and any extra electron would be placed in unoccupied states of other spin-channel. In this paper we use the convention that if N$_{\text{V}}> 24$, the spin channel where the half-metallic gap is, will be considered the minority spin channel. Accordingly, from Fig. 1, Supplementary material, it is clear that position of $E_F$ in the minority spin-channel would lead to the half-metals following different Slater-Pauling rules. If $E_F$ is situated between $t_{1u}$ and $t_{2g}$ states, it would follow M = $| \text{N}_{\text{V}} - 18 |$ rule. If the position of Fermi-energy is between $t_{1u}$ and $e_u$ states, the compound would follow M = $| \text{N}_{\text{V}} - 24 |$ rule. Simple difference in electron count in both spin-channel leads us to this conclusion.

\begin{figure}[h!]
\centering
\subfigure[]{\includegraphics[scale=0.2]{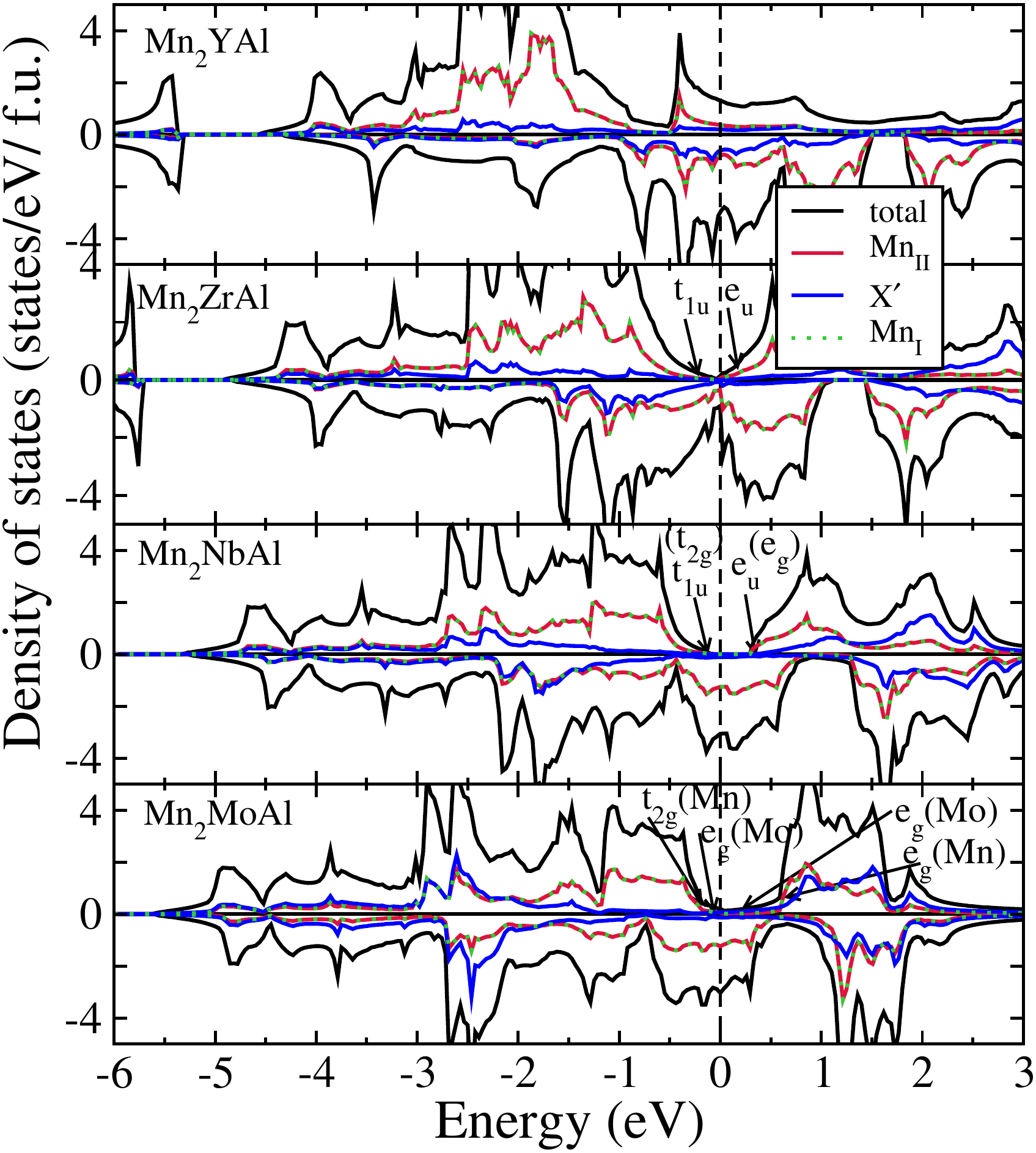}}
\subfigure[]{\includegraphics[scale=0.2]{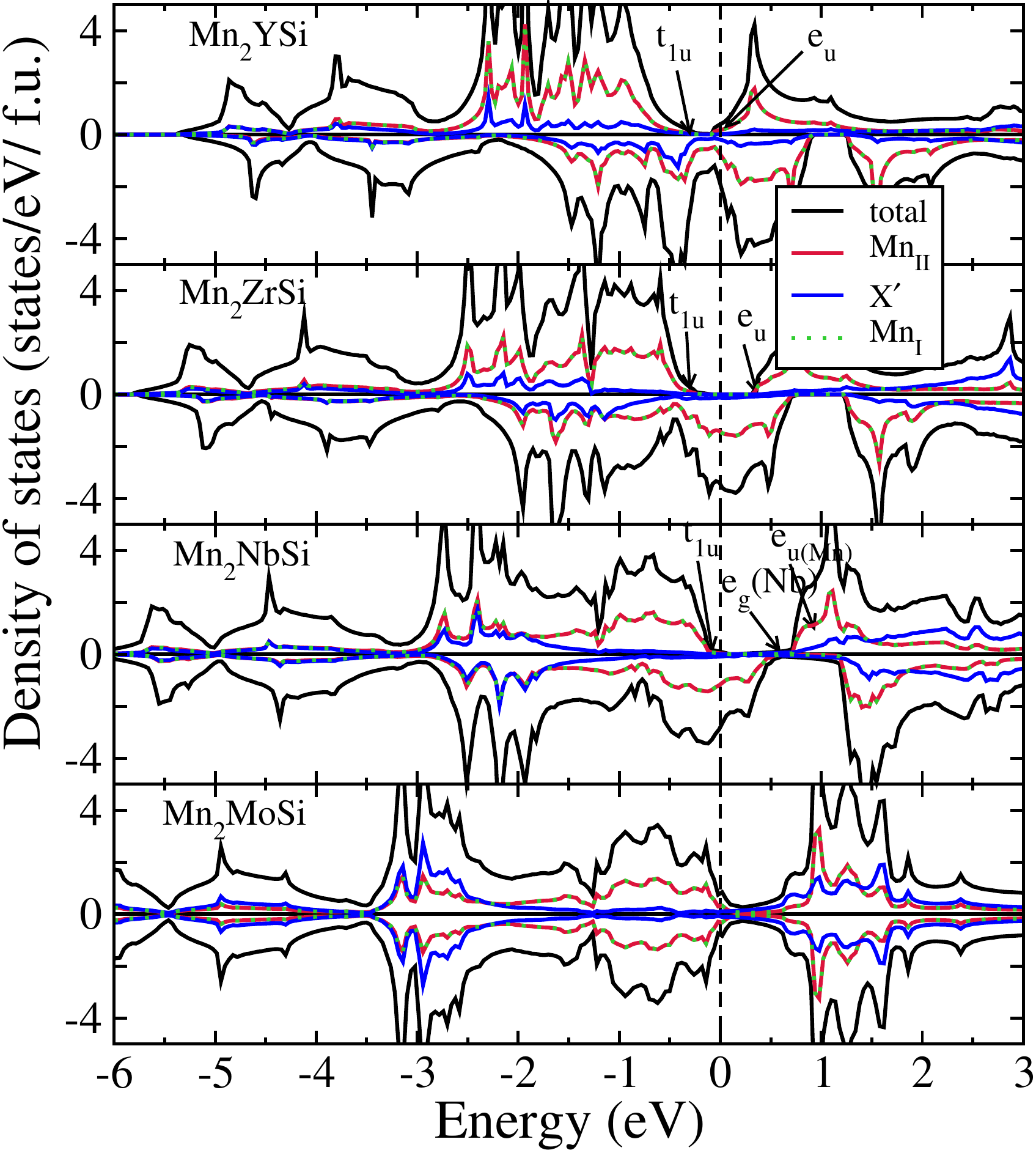}}
\caption{Spin polarized total and atom-projected densities of states for ({\bf a}) Mn$_2$X$^\prime$Al and ({\bf b}) Mn$_2$X$^\prime$Si ($\text{X}^\prime$= Y, Zr, Nb, Mo) compounds. The ground states of these compounds are Type-I. }
\label{DOSMn2YZ-t1}
\end{figure}

\begin{figure}[h!]
\centering
\subfigure[]{\includegraphics[scale=0.2]{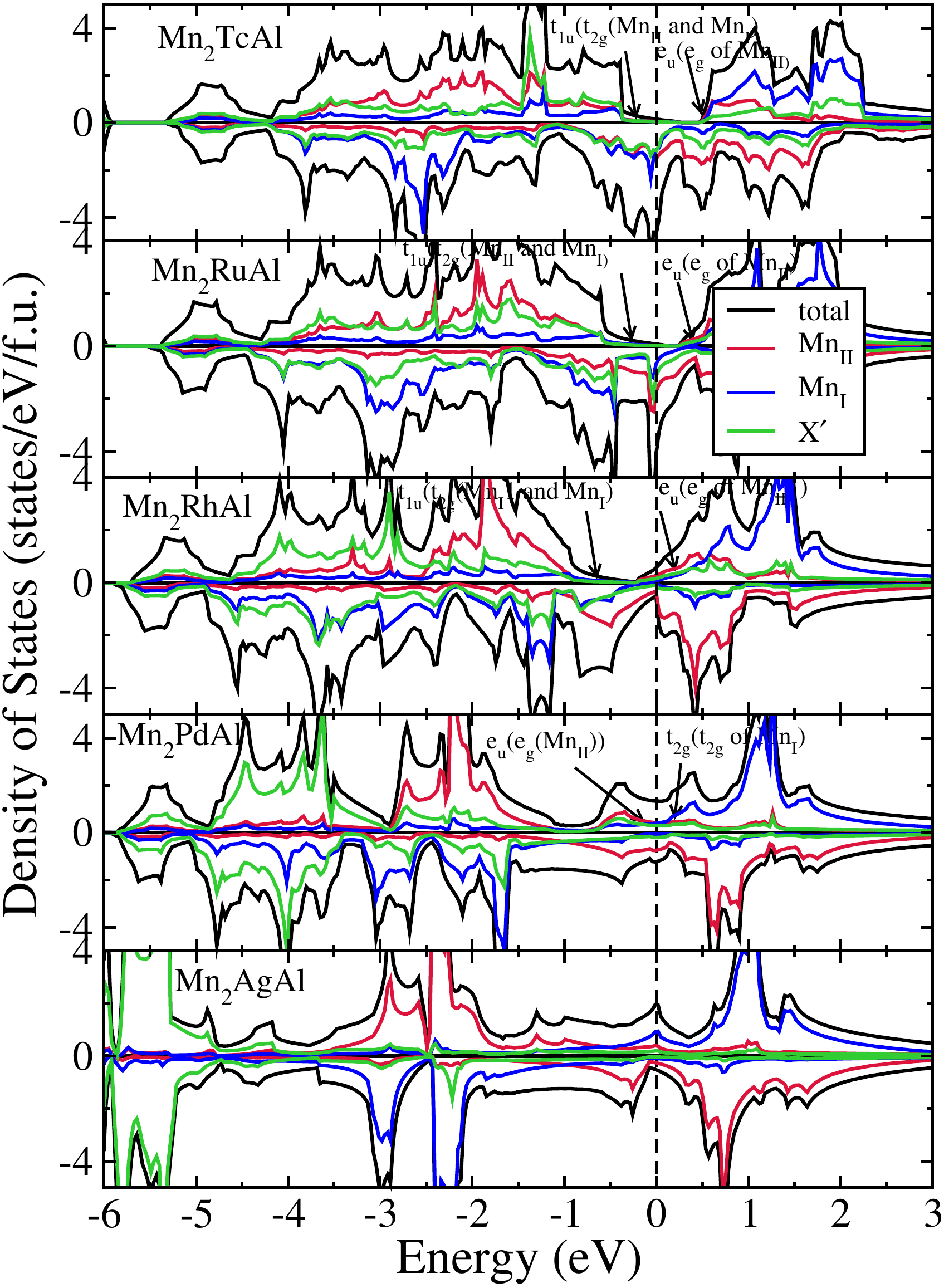}}
\subfigure[]{\includegraphics[scale=0.2]{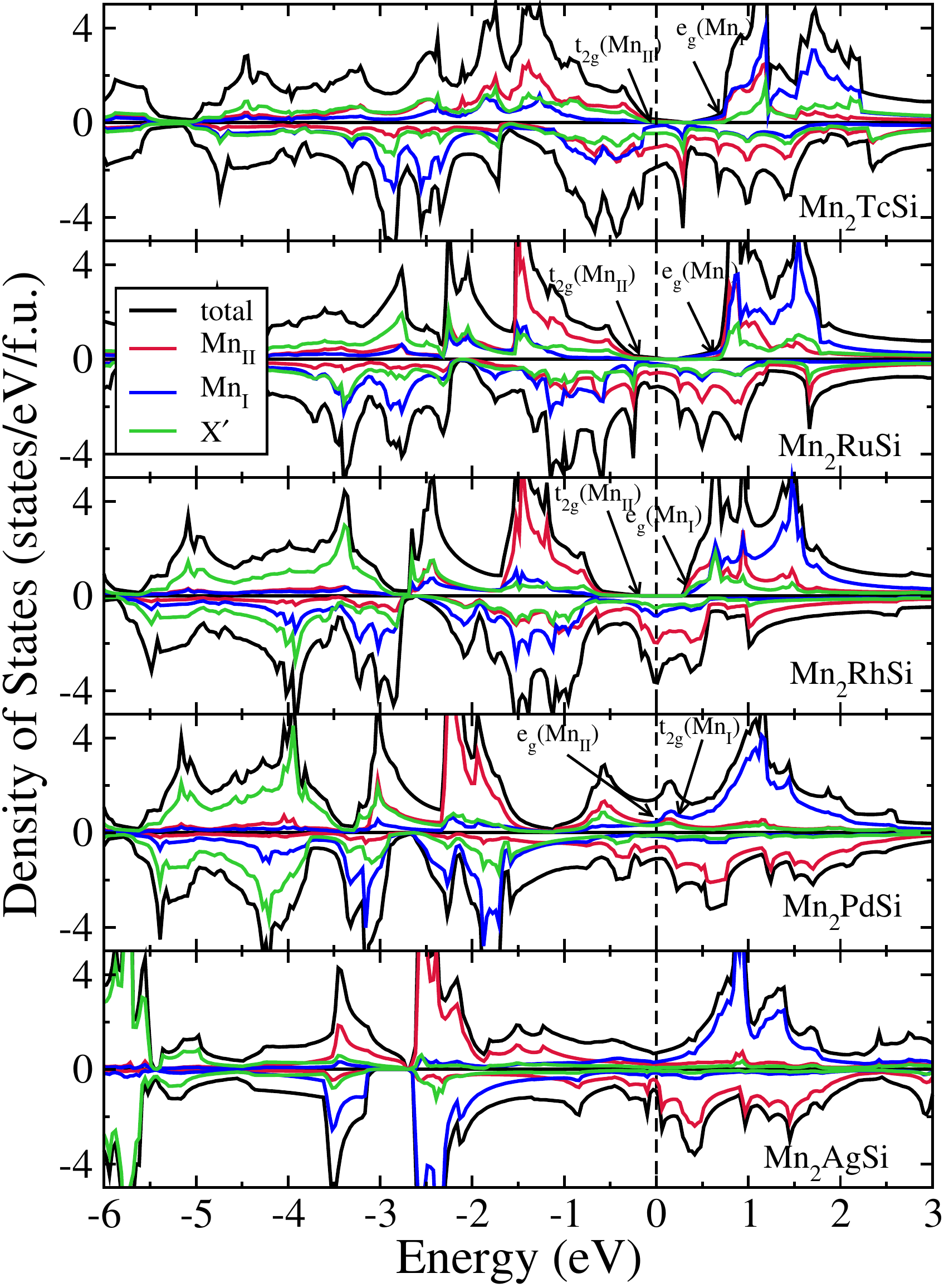}}
\caption{Spin polarized total and atom-projected densities of states for ({\bf a}) Mn$_2$X$^\prime$Al and ({\bf b}) Mn$_2$X$^\prime$Si ($\text{X}^\prime$= Y, Zr, Nb, Mo) compounds. The ground states of these compounds are Type-II. }
\label{DOSMn2YZ-t2}
\end{figure}

\begin{figure}[h!]
\centering
\subfigure[]{\includegraphics[scale=0.2]{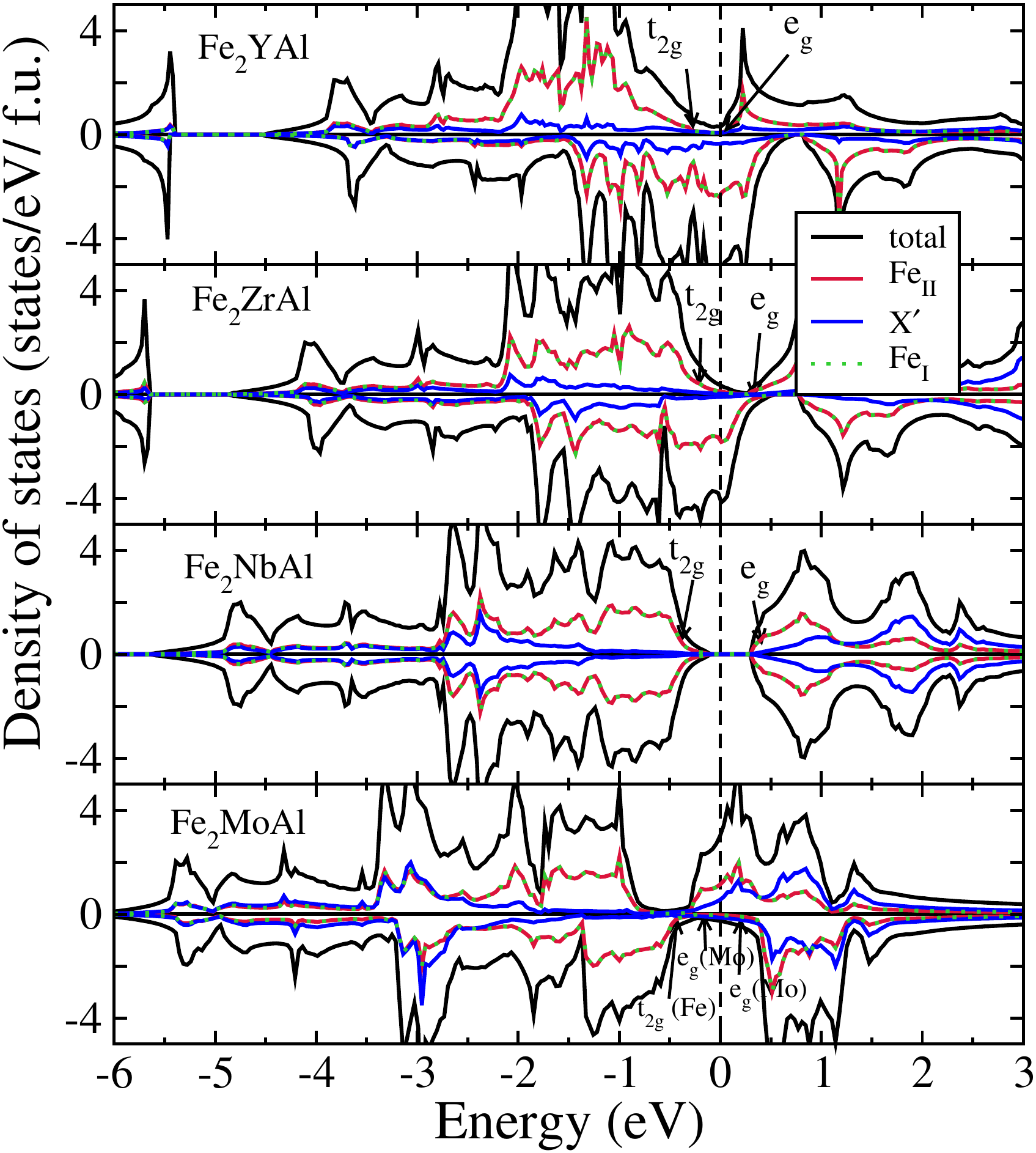}}
\subfigure[]{\includegraphics[scale=0.2]{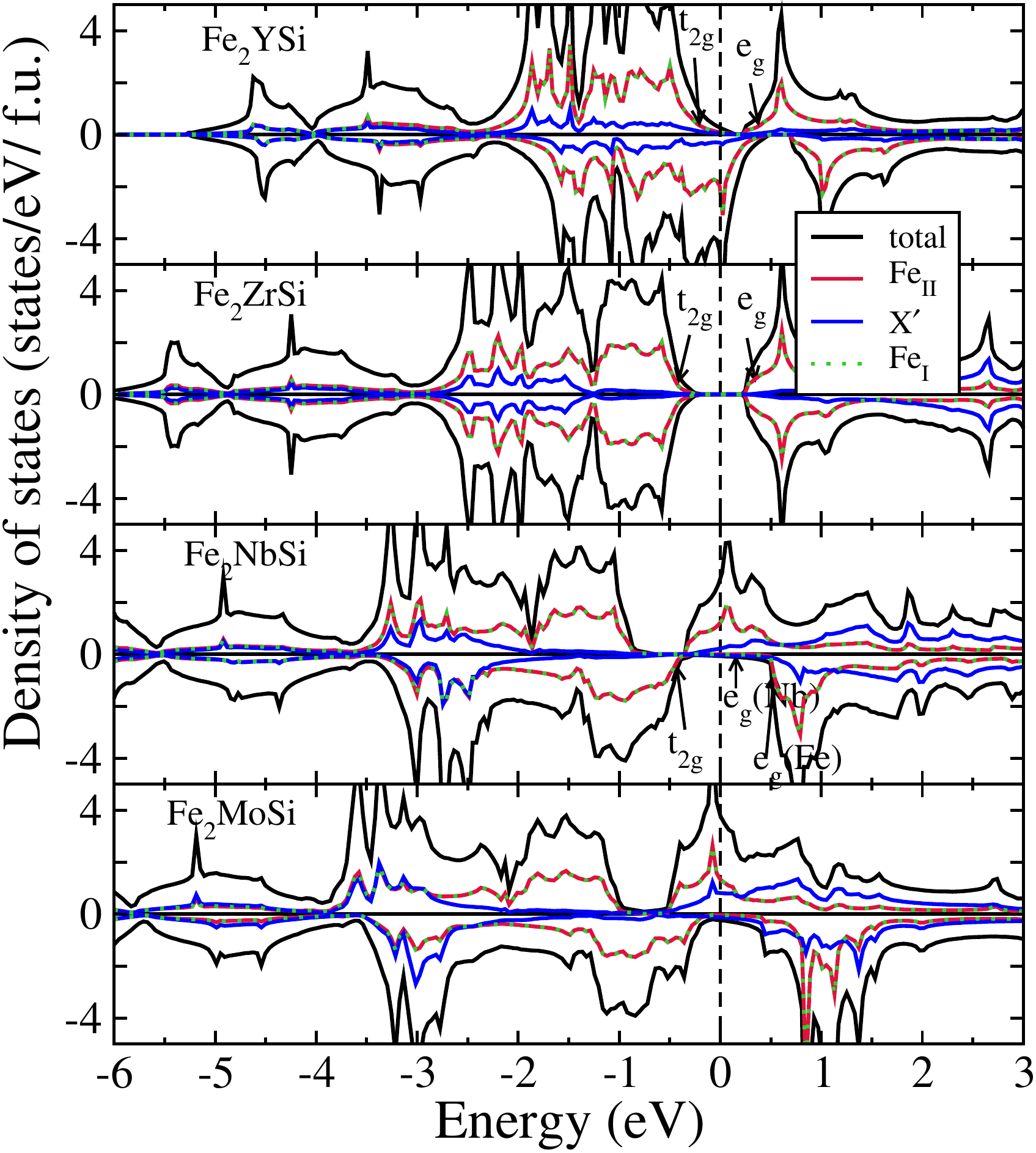}}
\caption{Spin polarized total and atom-projected densities of states for ({\bf a}) Fe$_2$X$^\prime$Al and ({\bf b}) Fe$_2$X$^\prime$Si ($\text{X}^\prime$= Y, Zr, Nb, Mo) compounds. The ground states of these compounds are Type-I. }
\label{DOSFe2YZ-t1}
\end{figure}

\begin{figure}[h!]
\centering
\subfigure[]{\includegraphics[scale=0.2]{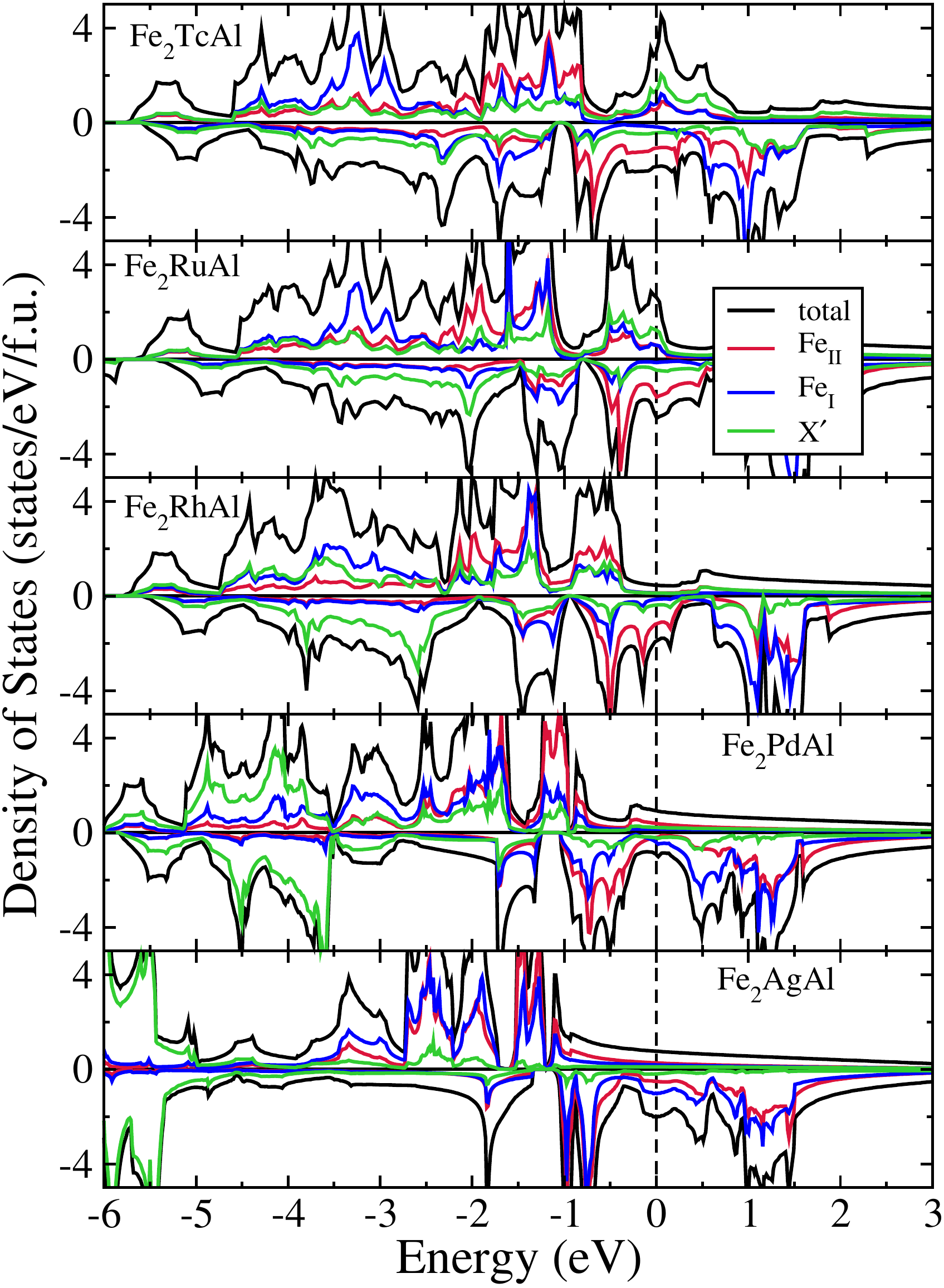}}
\subfigure[]{\includegraphics[scale=0.2]{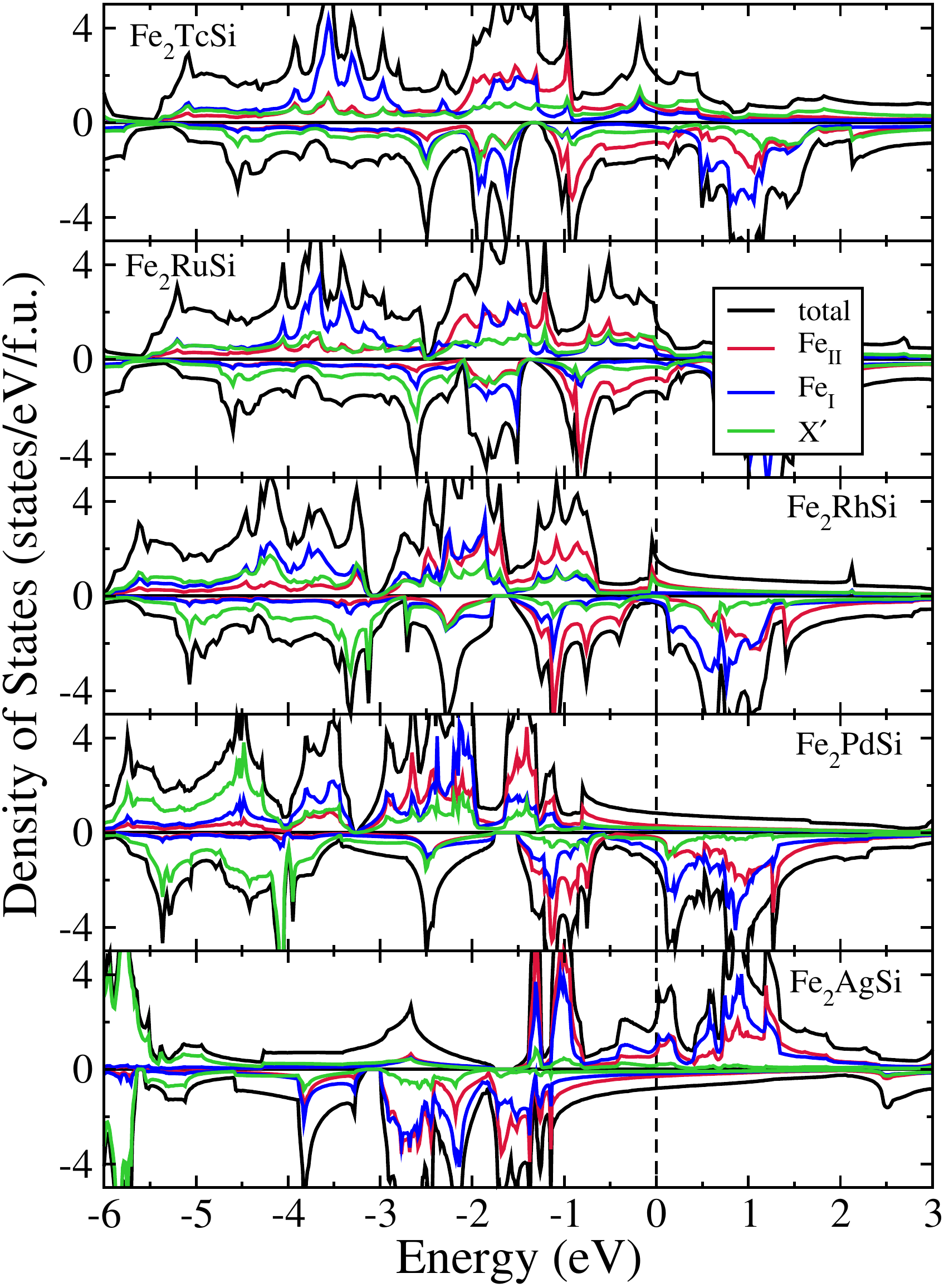}}
\caption{Spin polarized total and atom-projected densities of states for ({\bf a}) Fe$_2$X$^\prime$Al and ({\bf b}) Fe$_2$X$^\prime$Si ($\text{X}^\prime$= Y, Zr, Nb, Mo) compounds. The ground states of these compounds are Type-II. }
\label{DOSFe2YZ-t2}
\end{figure}

\begin{figure}[h!]
\centering
\subfigure[]{\includegraphics[scale=0.2]{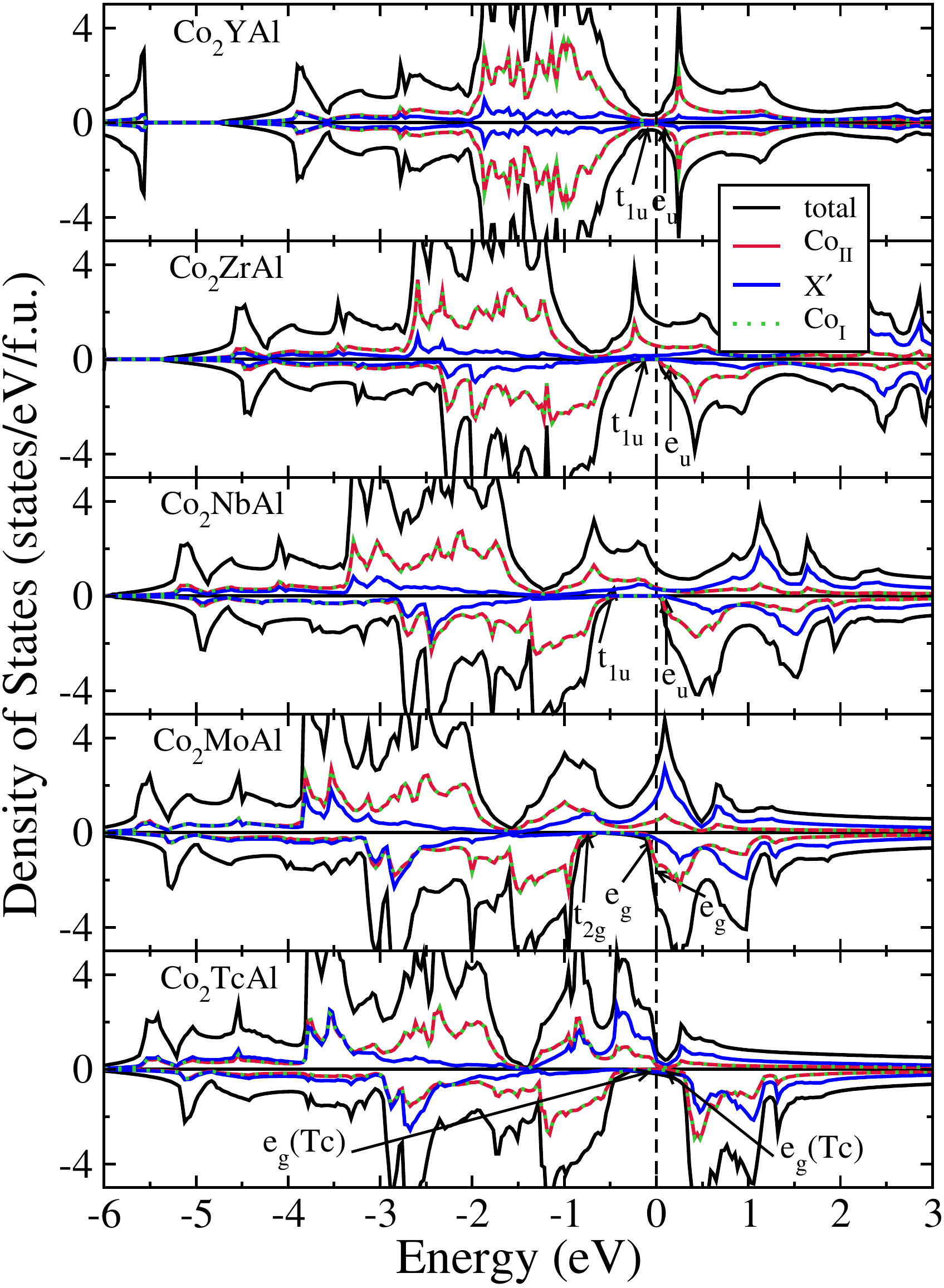}}
\subfigure[]{\includegraphics[scale=0.2]{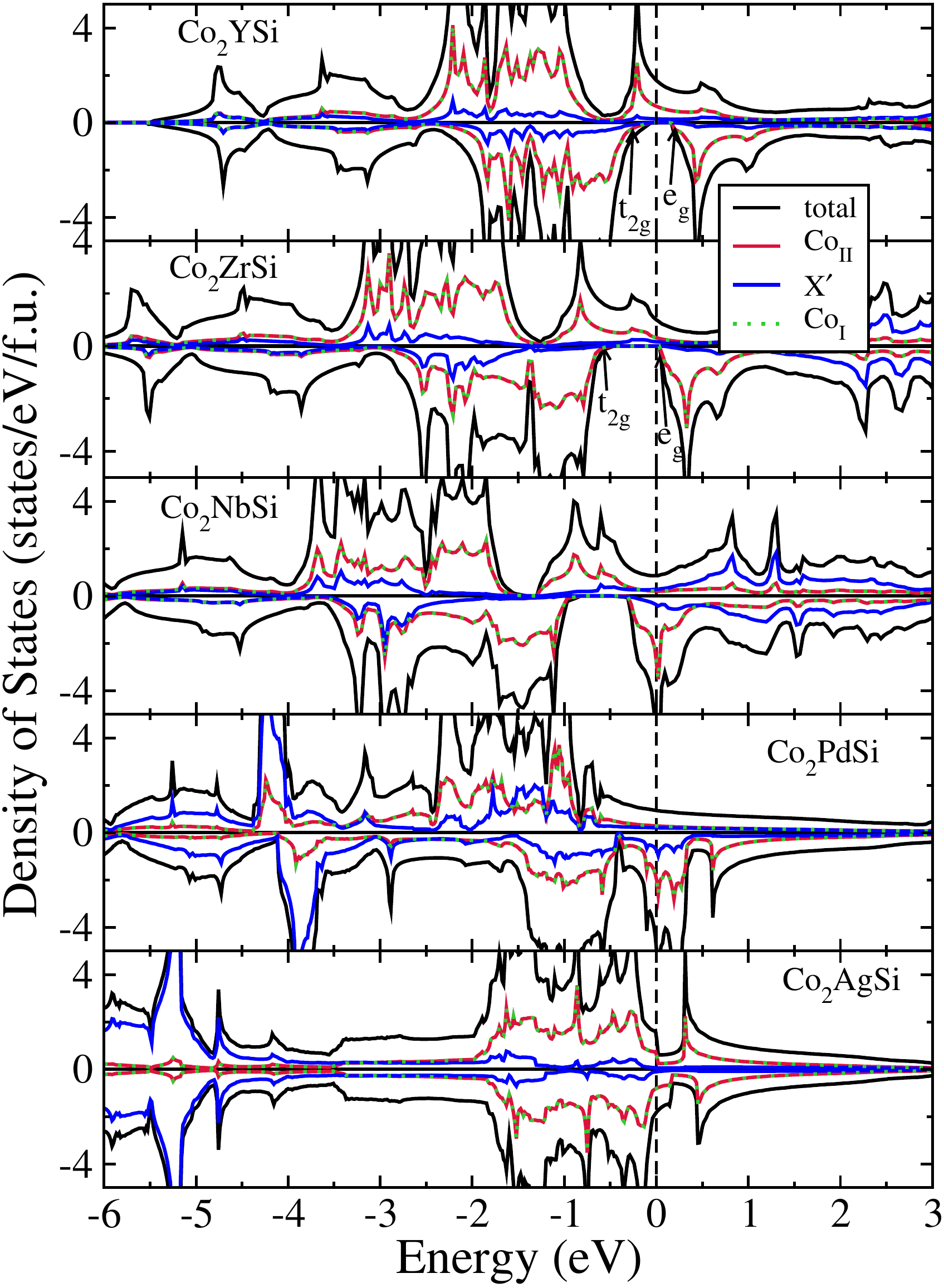}}
\caption{Spin polarized total and atom-projected densities of states for ({\bf a}) Co$_2$X$^\prime$Al and ({\bf b}) Co$_2$X$^\prime$Si ($\text{X}^\prime$= Y, Zr, Nb, Mo) compounds. The ground states of these compounds are Type-I. }
\label{DOSCo2YZ-t1}
\end{figure}

\begin{figure}[h!]
\centering
\subfigure[]{\includegraphics[scale=0.2]{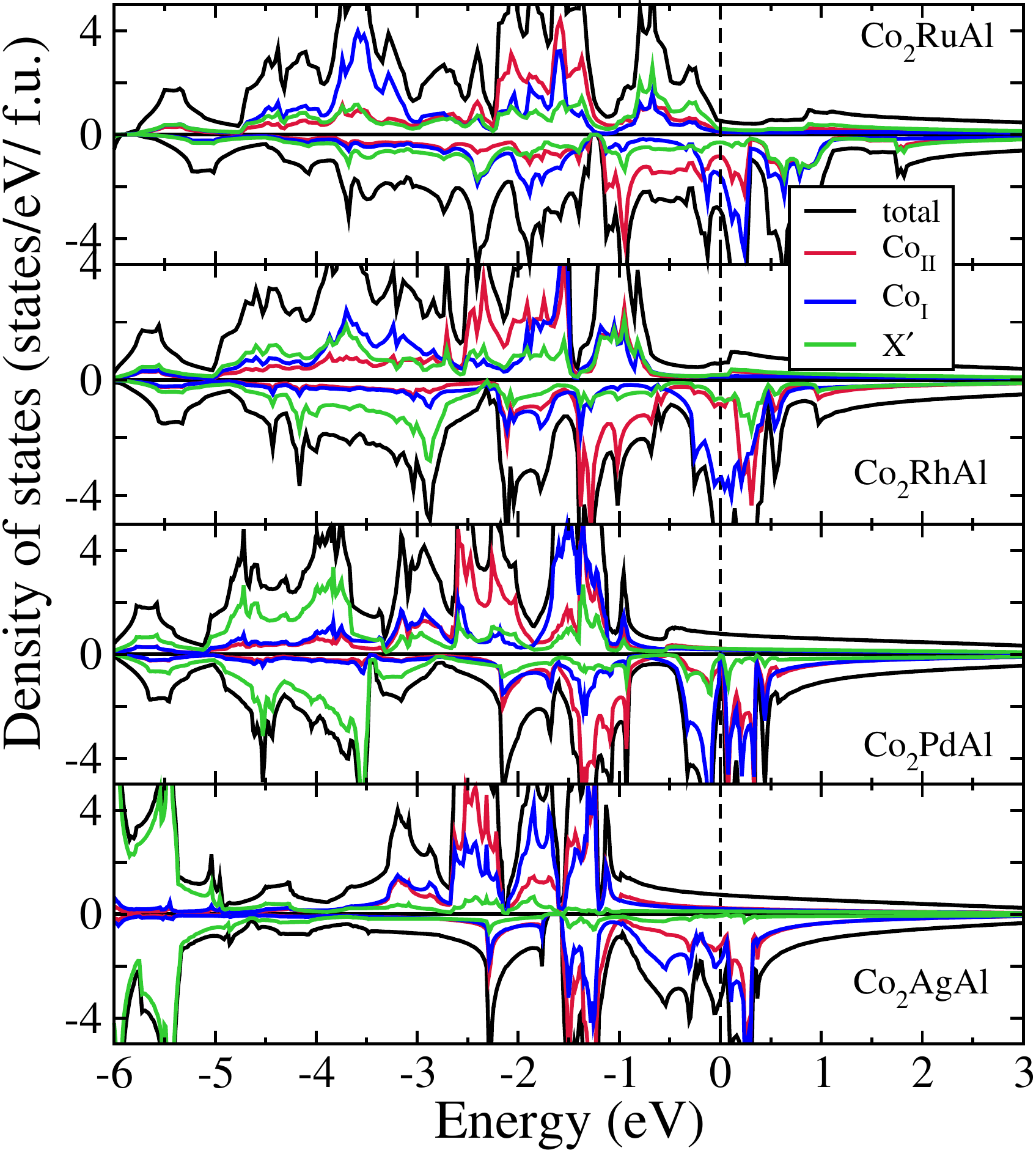}}
\subfigure[]{\includegraphics[scale=0.2]{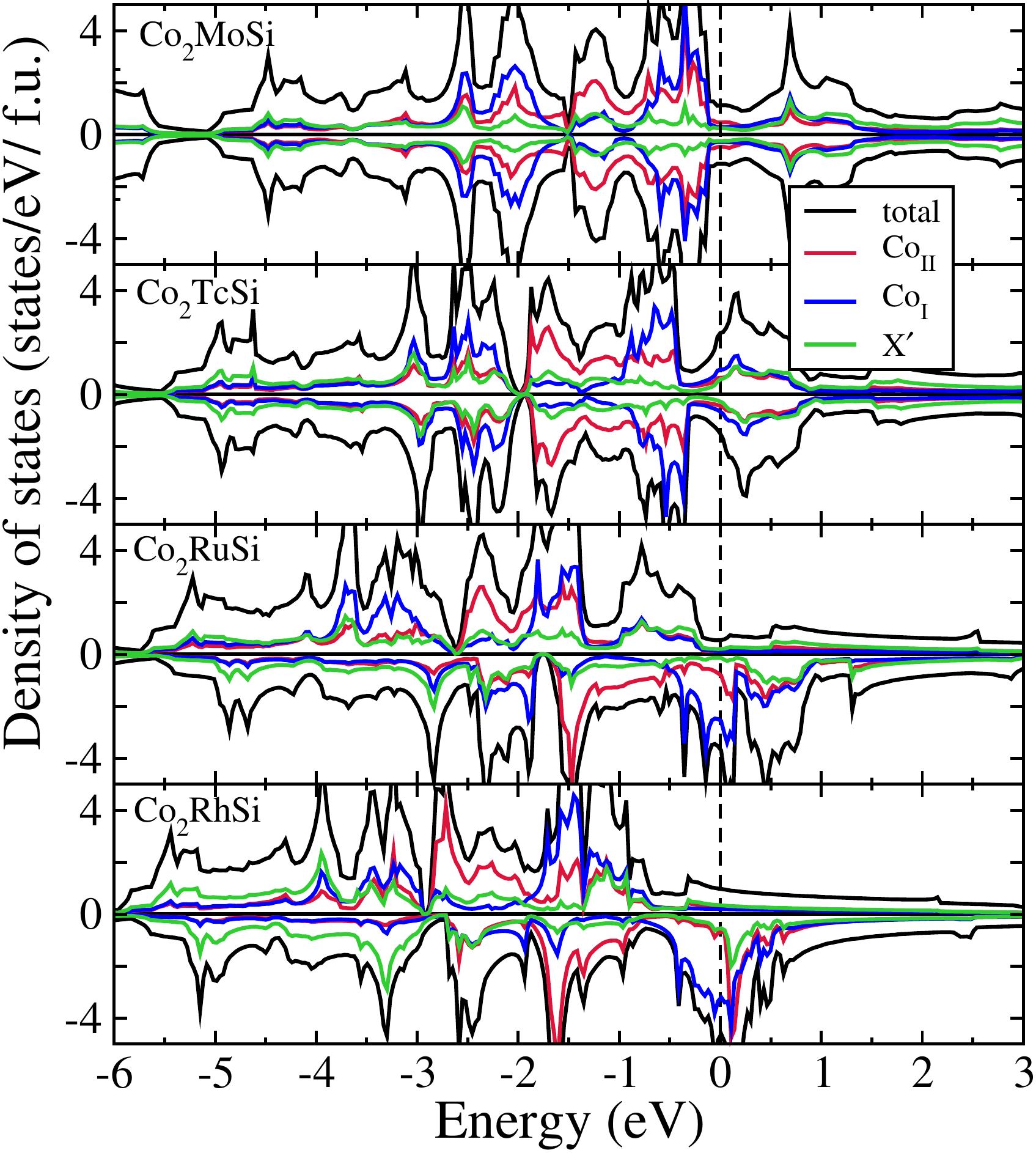}}
\caption{Spin polarized total and atom-projected densities of states for ({\bf a}) Co$_2$X$^\prime$Al and ({\bf b}) Co$_2$X$^\prime$Si ($\text{X}^\prime$= Y, Zr, Nb, Mo) compounds. The ground states of these compounds are Type-II. }
\label{DOSCo2YZ-t2}
\end{figure}

In Figs. \ref{DOSMn2YZ-t1}- \ref{DOSCo2YZ-t2} we present the spin-polarised total and atom projected densities of states of all X$_2$X$^\prime$Z compounds considered in this work  in their respective ground state structures. In the following we discuss them series-wise.

\subparagraph{\bf a.\quad Mn$_2$X$^\prime$Z compounds\\\\} 

The densities of states for compounds of Mn$_2$X$^\prime$Al (Mn$_2$X$^\prime$Si) series with T$_{\text{I}}$ and T$_{\text{II}}$ structures are shown, respectively, In Fig \ref{DOSMn2YZ-t1}.(a) and Fig \ref{DOSMn2YZ-t2}.(a) (Fig \ref{DOSMn2YZ-t1}.(b) and Fig \ref{DOSMn2YZ-t2}.(b)). We find several features common to either the compounds with same structure type or even across the structure types:  (a) for compounds in T$_{\text{I}}$ structure and with possible half-metallic gaps, the gaps are flanked by Mn $d$-states, while for compounds in T$_{\text{II}}$ structure, the contributions from Mn and X$^{\prime}$ are significant; both consistent with the generalised hybridisation picture described above, (b) across the structures, as the atomic number increases, the contributions from X$^{\prime}$ move gradually towards lower energies, (c) as long as structure types are same, the features in the densities of states of compounds from the two different series are very similar when N$_{\text{V}}$ are same. This is reflected more in compounds with structure T$_{\text{I}}$.  For example, Mn$_2$NbAl and Mn$_2$ZrSi both are half-metallic with N$_{\text{V}}$=22. Mn$_2$ZrAl and Mn$_2$YSi, Mn$_2$MoAl and Mn$_2$NbSi too have similar features in their densities of states. On the other hand, Mn$_2$MoSi and Mn$_2$TcAl, both are with N$_{\text{V}}$=24, yet have completely different features in their densities of states as the ground state structures are different.

In order to identify half-metals and understand their origin as N$_{\text{V}}$ changes continuously, we now focus on the densities of states of the compounds in the two series systematically. We find that for compounds crystallising in T$_{\text{I}}$ the origin of the gap in majority spin channel is due to the splitting of non-bonding $t_{1u}$ and $e_u$ states arising out of  hybridisations between the Mn atoms. For Mn$_2$ZrAl and Mn$_2$YSi the separation between $t_{1u}$ and $e_u$ states are not enough to extend the gap cutting through the Fermi-level to make these compounds half-metallic. Thus, inspite of their magnetic moments being integer or near integer, these compounds are not half-metals.
As we go from Mn$_2$ZrAl(Mn$_2$YSi) to Mn$_2$NbAl(Mn$_2$ZrSi), the extra electron is accommodated in one of the three $t_{1u}$ states in the majority band opening a half-metallic gap. 
In Mn$_2$MoAl (Mn$_2$NbSi) the extra electron is accommodated in one of the vacant t$_{1u}$ or in the e$_{g}$ states in the majority bands, thus producing high spin polarisation but destroying the half-metallicity as these states cut through the Fermi levels. Once again, inspite of having near integer moments, these two compounds at best can be considered near half-metals with Mn$_{2}$NbSi having a spin polarisation over 90$\%$.
Mn$_2$MoSi, with N$_{\text{V}}$=24, has same occupancy in both spin bands leading to zero magnetic moment, in conformation with the Slater-Pauling rule. However, inspite of Mn$_{2}$TcAl having N$_{\text{V}}$=24 too, the electronic structure is different from that of Mn$_{2}$MoSi as the former crystallises in T$_{\text{II}}$ structure. Though the moment is nearly zero in Mn$_{2}$TcAl (like Mn$_{2}$MoSi) it has a large spin polarisation of 94$\%$. This is because unlike Mn$_{2}$MoSi,  it has a near half-metallic gap in the majority spin band arising out of the separation of bonding t$_{2g}$ and non-bonding e$_{u}$ states across the Fermi level. This, in fact, is part of the general features observed in the electronic structure of the compounds with T$_{\text{II}}$ structure. In case of these compounds the bonding $t_{2g}$ hybrid orbitals, originating from all the transition metal atoms and the non-bonding $t_{1u}$ hybrids due to Mn and X$^\prime$ atoms lie extremely close; same happens for anti-bonding $e_g$ and non-bonding $e_u$ states. Throughout the Mn$_{2}$X$^{\prime}$Z series with T$_{\text{II}}$ structure, we find that the position of the gap and spin polarisation is determined by this. 

Thus, in Mn$_2$X$^\prime$Z series we found three true half-metals (Mn$_2$NbAl, Mn$_2$ZrSi and Mn$_2$RhSi), four near half-metals {\it i.e.} materials with high spin polarisation (greater than 90\%) (Mn$_2$TcAl, Mn$_2$RuAl, Mn$_2$NbSi, Mn$_2$RuSi). Out of the three true half-metals, Mn$_{2}$RhSi has also been found to be a potential thermoelectric material \cite{PKJhaPhysica18}. On the other hand, near half-metal Mn$_{2}$TcAl with it's high spin polarisation and near zero magnetic moment is a potential compensated ferrimagnet. It may be noted that Mn$_{3}$Al in DO$_{3}$ structure is a well known compensated ferrimagnet \cite{Mn3Al_PRA17} and that Mn$_{2}$TcAl is isoelectronic to it (derived by replacing one Mn with Tc), implying that the physical properties of magnetic Heusler compounds obtained by replacing one 3$d$ element with a 4$d$ may depend solely on N$_{\text{V}}$. We find a signature of this when we compare Mn$_{2}$YAl and Mn$_{2}$ScAl. Our calculations show that Mn$_{2}$YAl is not a half-metal at all, neither does it have a significant spin polarisation. However, it has exceptionally large magnetic moment exactly like Mn$_{2}$ScAl, which is isoelectronic.


\subparagraph{\bf b.\quad Fe$_2$X$^\prime$Z compounds\\\\} 
 
 Atom and spin resolved densities of states of Fe$_2$X$^\prime$Z  series are presented in the Fig \ref{DOSFe2YZ-t1} and Fig \ref{DOSFe2YZ-t2}. Fig \ref{DOSFe2YZ-t1}.(a) and Fig \ref{DOSFe2YZ-t1}.(b) (Fig \ref{DOSFe2YZ-t2}.(a) and Fig \ref{DOSFe2YZ-t2}.(b)) are the densities of states  for compounds with structure type T$_{\text{I}}$(T$_{\text{II}}$). We find substantial similarities between Fe$_{2}$X$^\prime$Z and Mn$_{2}$X$^{\prime}$Z as far as some of the general features in the electronic structures are concerned.  Like in Mn$_{2}$X$^{\prime}$Z series, the main contributions to densities of states for compounds in Fe$_{2}$X$^{\prime}$Z near Fermi levels come from the Fe atoms in case of compounds with structure T$_{\text{I}}$ whereas in case of compounds with T$_{\text{II}}$ structure, the main contributions come from the tetrahedrally co-ordinated Fe and X$^{\prime}$ atoms. We also find that with increasing atomic number the X$^\prime$ $d$- states lie deeper into the valence band, a feature similar to the Mn$_{2}$X$^{\prime}$Z series. The major difference, however, between the compounds in two series, is that none of the 18 compounds in Fe$_2$X$^\prime$ series is found to be half-metallic. This difference originates from the fact that the Fe states are more delocalised as compared to Mn states. As a result, often the Fe states or the ones with hybridisations with Fe atoms, extend into the unoccupied part, leaving no possibility of opening up of a half-metallic gap. However, in almost all compounds, there is either a pseudo gap in the occupied part of one spin channel and a gap like valley cutting through the Fermi level in the other. The trends in the features of the electronic structures as one goes through the compounds in a series and with a given crystal structure, are same across the Al and Si series. The electronic structures are very similar if N$_{\text{V}}$ is same. For compounds with structure type T$_{\text{I}}$, the electronic structure near the Fermi level is contributed mostly by the t$_{2g}$ and e$_{g}$ Fe states in occupied and unoccupied parts respectively. We find the only near half-metal in Fe-series compounds to be Fe$_{2}$NbSi with spin polarisation 95$\%$ crystallising in T$_{\text{I}}$ structure. The compound right before Fe$_{2}$NbSi in the series is Fe$_{2}$ZrSi which has 24 valence electrons and turns out to be a non-magnetic semiconductor. It is interesting to note that unlike Mn$_{2}$X$^{\prime}$Z, the changes in the electronic structure of Fe$_{2}$X$^{\prime}$Z compounds, when N$_{\text{V}}$ changes from 23 to 24 are drastic. In the Mn-series, the changes in electronic structures between Mn$_{2}$NbAl (N$_{\text{V}}$=23) and Mn$_{2}$MoAl (N$_{\text{V}}$=24) bore a continuity while the changes from Fe$_{2}$ZrAl (Fe$_{2}$YSi) to Fe$_{2}$NbAl (Fe$_{2}$ZrSi) are sharp and substantial. The changes from Fe$_{2}$NbAl (Fe$_{2}$ZrSi) to Fe$_{2}$MoAl (Fe$_{2}$NbSi) again are systematic, following a trend. The extra electron in Fe$_{2}$MoAl (in comparison to Fe$_{2}$ZrAl) is accommodated in the spin-up band, which reflects in a pseudo gap inside the occupied part of spin up channel. The semiconducting gap of Fe$_{2}$NbAl in the spin down channel would have been intact if the hybridisations near the Fermi level wouldn't have changed upon replacement of Nb with Mo. The hybridisations of Mo and Fe states near the Fermi level produce a valley cutting through the Fermi level, reducing the magnetic moment significantly from the Slater-Pauling predicted value, bringing down the spin polarisation as well. The semiconducting gap of spin down channel in Fe$_{2}$ZrSi nearly survives in Fe$_{2}$NbSi as the Nb e$_{g}$ and Fe t$_{2g}$ hybridisations near Fermi level are weak as compared to Fe$_{2}$MoAl. For the compounds with structure type T$_{\text{II}}$, the proximity of $d$ states of Fe and X$^{\prime}$ mix the states substantially in both spin channels, leaving little possibility of a semiconducting gap in any of the spin channels.

\subparagraph{\bf c.\quad Co$_2$X$^\prime$Z compounds\\\\} 

In Figs \ref{DOSCo2YZ-t1} - \ref{DOSCo2YZ-t2} we show the total and partial densities of states of Co$_2$X$^\prime$Z compounds. Like observed in cases of Mn$_{2}$ and Fe$_{2}$ based compounds, the Co$_{2}$ based compounds with structure T$_{\text{I}}$ have Co states dominating the features near the half-metallic gaps for half-metals. We find more half-metals with 100$\%$ spin polarisations among Co$_{2}$X$^{\prime}$Z compounds in comparison to compounds in other two series. All the half-metals in Co$_{2}$-compounds are with T$_{\text{I}}$ structure, arising because of the Co states, localised more in comparison to Mn or Fe in Mn$_{2}$ and Fe$_{2}$ based compounds respectively. This feature is also seen in the more familiar Co$_{2}$X$^{\prime}$Z compounds with X$^{\prime}$ a 3$d$ element \cite{co2ysi_3d23dsi,GalanakisPRB02,Co2yZ_3d23dz}. The Co-based compounds studied in the present work have N$_{\text{V}}$ greater than or equal to 24. Thus the half-metallic gaps are found in the spin down channels. In Fig \ref{DOSCo2YZ-t1}.(a) and Fig \ref{DOSCo2YZ-t1}.(b) we see that the main contributions to densities of states comes from Co-$t_{2g}$ or Co-$t_{1u}$ below the Fermi-level and from Co-$e_g$ or Co-$e_u$ above the Fermi-level in both Co$_2$X$^\prime$Al and Co$_2$X$^\prime$Si series. Co$_2$YAl, having N$_{\text{V}}$=24, is a non-magnetic compound where each spin band has exactly 12 electrons. In Co$_2$ZrAl -Co$_{2}$NbAl (Co$_2$YSi-Co$_2$ZrSi) minority spin channel have 12 electrons and the extra electrons are fully accommodated in the majority spin channel opening the gap in minority spin channel. In  Co$_2$MoAl (Co$_2$NbSi) the extra electron is not fully accommodated in the next $e_g$ band in spin up channel, instead it is shared by both the spin channels, thus destroying the half-metallicity. In Co$_2$TcAl the extra electron, as expected, is not placed totally in the anti-bonding $e_g$ states in spin up channel and there is considerable mixing of Co and Tc e$_{g}$ states near Fermi level leaving little chance of half-metallicity. Co$_{2}$PdSi and Co$_{2}$AgSi, the other two compounds with T$_{\text{I}}$ structure have different hybridisation pattern where delocalised Pd and Ag states hybridise considerably with Co states near Fermi level, leading them to behave like normal metals. 

The compounds with T$_{\text{II}}$ structures have significant hybridisations between Co and X$^{\prime}$ states in both spin channels and thus there is no possible half-metals with this crystal structure. Like the Mn$_{2}$ and Fe$_{2}$ series, this is an artefact of the atomic arrangement in T$_{\text{II}}$ structure. Thus, in Co$_{2}$X$^{\prime}$Z series we find four half-metals, Co$_{2}$ZrAl, Co$_{2}$NbAl, Co$_{2}$ZrSi and Co$_{2}$NbSi, having 100$\%$ spin polarisation.


\subsection{Trends in the local magnetic moments}
In Tables \ref{table3} to \ref{table8} we present the total and site projected magnetic moments along with the spin polarisations for all the compounds considered here. We try to understand the magnetic structures, the contributions of each transition metal atom towards the total moments, and the trends in variations of the atomic moments across series and structures. We also intend to relate such trends with the trends in the electronic structures discussed in previous sub-section.

The 3$d$ elements contribute most towards the magnetic structure and the overall magnetisation in all the compounds studied here. This is a reflection of the facts that the electronic structures of these compounds evolve around the electronic structures of the 3$d$ constituents throughout the series. For compounds with structure type T$_{\text{I}}$ we find that the moments of the 3$d$ constituents are very close across the corresponding Al and Si series, the only exceptions being Co$_{2}$MoAl and Co$_{2}$NbSi. For compounds with structure type T$_{\text{II}}$, there is no such trend. For Mn$_{2}$-compounds, Mn moments differ substantially between the compounds in Al series and those in Si series. The moments of Mn$_{\text{I}}$, the Mn atoms at the 4c sites, are considerably reduced in the compounds of Si series, as compared to those in the Al series. The trend is not so in cases of the Fe- and Co-series, except few cases like Fe$_{2}$RuAl-Fe$_{2}$TcSi and Co$_{2}$RuAl-Co$_{2}$TcSi pairs.

The moments of Mn atoms in both Mn$_{2}$X$^{\prime}$Al and Mn$_{2}$X$^{\prime}$Si series decrease with N$_{\text{V}}$ for compounds with structure type T$_{\text{I}}$. The opening of the gap in the spin up band and subsequent increase in the electrons in the spin down band, as explained in the previous sub-section, is responsible for such trend. For compounds with structure type T$_{\text{II}}$, the two Mn atoms are ferrimagnetically coupled due to they being nearest neighbours. We find that in both Al- and Si-series, the moment of Mn$_{\text{II}}$, the Mn atom at the 4a sites, is much robust and increases gradually with N$_{\text{V}}$. This is because of the fact that the Mn$_{\text{II}}$ spin up band is nearly full with a near gap close to the Fermi level while the spin down band gradually becomes empty. The Mn$_{\text{I}}$ atom, on the other hand, hybridises well with X$^{\prime}$ as they occupy symmetric positions and thus loses the robustness in it's moment. The re-distributions of states among the spin bands due to the hybridisations of Mn and X$^{\prime}$ $d$-orbitals are more prominent in the compounds of the Si-series which is responsible for the substantial reductions in the Mn$_{\text{I}}$ moments.

The drastic changes in the electronic structures of Fe$_{2}$-compounds crystallising in T$_{\text{I}}$ structure when N$_{\text{V}}$ is equal to or greater than 24 reflect in the Fe moments. The Fe moment decreases significantly from Fe$_{2}$YAl to Fe$_{2}$ZrAl. The extra electron in Fe$_{2}$ZrAl occupies the spin down band of Fe reducing the moment. The moment starts to pick up after N$_{\text{V}}$ is beyond 24. The extra electron now starts to occupy the Fe spin up band primarily. This explains the trends in the Fe$_{2}$X$^{\prime}$Si compounds. Like observed in the Mn-series, Fe$_{\text{II}}$ moment in compounds with structure type T$_{\text{II}}$ is robust. Although the magnetic interaction between Fe atoms is ferromagnetic, substantial states in Fe$_{\text{I}}$ close to Fermi level , an artefact of hybridisations with X$^{\prime}$, reduces it's moment in comparison to Fe$_{\text{II}}$.

In contrast, Co moment in Co$_{2}$X$^{\prime}$Z compounds with structure type T$_{\text{I}}$ increases with N$_{\text{V}}$, the only exception being Co$_{2}$NbSi where the Co moment decreases in comparison to that in Co$_{2}$ZrSi. The electronic structures provide clue to this trend. As the extra electrons available with increasing N$_{\text{V}}$ are gradually accommodated primarily in the spin up bands, the moment increases. In case of Co$_{2}$NbSi, the extra electron available with respect to Co$_{2}$ZrSi is shared between both spin bands, thus departing from the general trend. Same trend is seen Co$_{\text{II}}$ moment in compounds with structure type T$_{\text{II}}$. The Co$_{\text{I}}$ moments, though reduced in comparison to those of Co$_{\text{II}}$, the reduction is less substantial in comparison to Mn$_{2}$- and Fe$_{2}$-compounds. The extra electron available as N$_{\text{V}}$ increases is accommodated primarily in the spin down band, reducing the moment of Co$_{\text{II}}$. The moment of Co$_{\text{I}}$, on the other hand, hardly changes from compound to compound. 

In all the compounds, the X$^{\prime}$ atom contributes to the overall magnetic moment, mostly for structure type T$_{\text{II}}$. The greater hybridisation with the 3$d$ element, as a consequence of geometry, is responsible for this. Al and Si atoms have vanishingly small contributions in all cases.
\begin{table}
\caption{\label{table3}Total and atomic magnetic moment of Mn$_{2}$X$^{\prime}$Al systems in $\mu_{B}/f.u.$.  N$_{\text{V}}$ is the number of valence electron of the systems. M is the total moment and M$_{\text{i}}$ is the moment of constituent $\text{i}$.}
\vspace{2 mm}
\centering
\begin{tabular}{ c c c c c c c c c c}
\hline\hline
Systems(T$_\text{I}$) &  N$_{\text{V}}$ & M     & M$_{\text{Mn$_{\text{I}}$}}$ & M$_{\text{Mn$_{\text{II}}$}}$ & M$_{X^{\prime}}$ & M$_{Al}$  & P($\%$)  \\ [0.1ex]
\hline
Mn$_2$YAl &         20 &    5.72 &          2.99 &          2.99 &          -0.13 &         -0.07 &         39 \\
Mn$_2$ZrAl &        21 &    3.00 &          1.73 &          1.73 &          -0.34 &         -0.05 &         74 \\
Mn$_2$NbAl &        22 &    2.00 &          1.21 &          1.21 &          -0.36 &         -0.03 &         100 \\
Mn$_2$MoAl &        23 &    1.04 &          0.68 &          0.68 &          -0.31 &         -0.01 &         88 \\[1ex]
\hline
Systems(T$_\text{II}$)& N$_{\text{V}}$ & M     & M$_{\text{Mn$_{\text{I}}$}}$ & M$_{X^{\prime}}$ & M$_{\text{Mn$_{\text{II}}$}}$ &   M$_{Al}$  & P($\%$)   \\ [0.1ex]
\hline
Mn$_2$TcAl &        24 &    0.04 &          -2.35 &         -0.37 &         2.75 &          0.03 &          94 \\
Mn$_2$RuAl &        25 &    1.01 &          -2.18 &         0.06 &          3.07 &          0.02 &          96 \\
Mn$_2$RhAl &        26 &    1.86 &          -1.84 &         0.34 &          3.26 &          0.01 &          6 \\
Mn$_2$PdAl &        27 &    0.86 &          -2.80 &         0.13 &          3.45 &          0.01 &          4 \\
Mn$_2$AgAl &        28 &    0.31 &          -3.12 &         0.03 &          3.40 &          0.01 &          59 \\ [1ex]

\hline\hline
\end{tabular}
\end{table}

\begin{table}
\caption{\label{table4}otal and atomic magnetic moment of Mn$_{2}$X$^{\prime}$Si systems in $\mu_{B}/f.u.$.  N$_{\text{V}}$ is the number of valence electron of the systems. M is the total moment and M$_{\text{i}}$ is the moment of constituent $\text{i}$..}
\vspace{2 mm}
\centering
\begin{tabular}{ c c c c c c c c c c}
\hline\hline
Systems(T$_\text{I}$) &  N$_{\text{V}}$ & M     & M$_{\text{Mn$_{\text{I}}$}}$ & M$_{\text{Mn$_{\text{II}}$}}$ & M$_{X^{\prime}}$ & M$_{Si}$  & P($\%$)  \\ [0.1ex]
\hline
Mn$_2$YSi & 	 21 & 	 3.02 & 	 1.64 & 	 1.64 & 	 -0.10 & 	 -0.07 & 	 66 \\ 
Mn$_2$ZrSi & 	 22 & 	 2.00 & 	 1.12 & 	 1.12 & 	 -0.20 & 	 -0.04 & 	 100 \\ 
Mn$_2$NbSi & 	 23 & 	 0.99 & 	 0.59 & 	 0.59 & 	 -0.17 & 	 -0.02 & 	 93 \\ 
Mn$_2$MoSi & 	 24 & 	 -0.00 & 	 -0.00 & 	 -0.00 & 	 0.00 & 	 0.00 & 	 0 \\ [1ex]
\hline
Systems(T$_\text{II}$)& N$_{\text{V}}$ & M     & M$_{\text{Mn$_{\text{I}}$}}$ & M$_{X^{\prime}}$ & M$_{\text{Mn$_{\text{II}}$}}$ &   M$_{Si}$  & P($\%$)   \\ [0.1ex]
\hline
Mn$_2$TcSi & 	 25 & 	 1.04 & 	 -1.38 & 	 -0.30 & 	 2.63 & 	 0.05 & 	 87 \\ 
Mn$_2$RuSi & 	 26 & 	 2.00 & 	 -0.85 & 	 0.07 & 	 2.70 & 	 0.03 & 	 91 \\ 
Mn$_2$RhSi & 	 27 & 	 3.00 & 	 -0.44 & 	 0.30 & 	 3.04 & 	 0.02 & 	 100 \\ 
Mn$_2$PdSi & 	 28 & 	 0.95 & 	 -2.59 & 	 0.08 & 	 3.37 & 	 0.04 & 	 13 \\ 
Mn$_2$AgSi & 	 29 & 	 0.59 & 	 -2.88 & 	 0.02 & 	 3.39 & 	 0.04 & 	 15 \\  [1ex]

\hline\hline
\end{tabular}
\end{table}
\begin{table}
\caption{\label{table5}Total and atomic magnetic moment of Fe$_{2}$X$^{\prime}$Al systems in $\mu_{B}/f.u.$.  N$_{\text{V}}$ is the number of valence electron of the systems. M is the total moment and M$_{\text{i}}$ is the moment of constituent $\text{i}$..}
\vspace{2 mm}
\centering
\begin{tabular}{ c c c c c c c c c c}
\hline\hline
Systems(T$_\text{I}$) &  N$_{\text{V}}$ & M     & M$_{\text{Fe$_{\text{I}}$}}$ & M$_{\text{Fe$_{\text{II}}$}}$ & M$_{X^{\prime}}$ & M$_{Al}$  & P($\%$)  \\ [0.1ex]
\hline
Fe$_2$YAl & 	 22 & 	 1.95 & 	 1.17 & 	 1.17 & 	 -0.18 & 	 -0.04 & 	 89 \\ 
Fe$_2$ZrAl & 	 23 & 	 0.91 & 	 0.54 & 	 0.54 & 	 -0.12 & 	 -0.01 & 	 85 \\ 
Fe$_2$NbAl & 	 24 & 	 0.00 & 	 0.00 & 	 0.00 & 	 0.00 & 	 0.00 & 	 0 \\ 
Fe$_2$MoAl & 	 25 & 	 0.82 & 	 0.48 & 	 0.48 & 	 -0.07 & 	 -0.01 & 	 84 \\ [1ex]
\hline
Systems(T$_\text{II}$)& N$_{\text{V}}$ & M     & M$_{\text{Fe$_{\text{I}}$}}$ &M$_{X^{\prime}}$ & M$_{\text{Fe$_{\text{II}}$}}$ &   M$_{Al}$  & P($\%$)   \\ [0.1ex]
\hline
Fe$_2$TcAl & 	 26 & 	 3.29 & 	 1.51 & 	 -0.26 & 	 2.06 & 	 -0.01 & 	 22 \\ 
Fe$_2$RuAl & 	 27 & 	 5.20 & 	 2.23 & 	 0.49 & 	 2.59 & 	 -0.03 & 	 2 \\ 
Fe$_2$RhAl & 	 28 & 	 5.00 & 	 1.98 & 	 0.39 & 	 2.77 & 	 -0.04 & 	 62 \\ 
Fe$_2$PdAl & 	 29 & 	 4.84 & 	 2.00 & 	 0.14 & 	 2.78 & 	 -0.04 & 	 1 \\ 
Fe$_2$AgAl & 	 30 & 	 4.75 & 	 2.19 & 	 -0.00 & 	 2.63 & 	 -0.04 & 	 44 \\  [1ex]

\hline\hline
\end{tabular}
\end{table}

\begin{table}
\caption{\label{table6}Total and atomic magnetic moment of Fe$_{2}$X$^{\prime}$Si systems in $\mu_{B}/f.u.$.  N$_{\text{V}}$ is the number of valence electron of the systems. M is the total moment and M$_{\text{i}}$ is the moment of constituent $\text{i}$..}
\vspace{2 mm}
\centering
\begin{tabular}{ c c c c c c c c c c}
\hline\hline
Systems(T$_\text{I}$) &  N$_{\text{V}}$ & M     & M$_{\text{Fe$_{\text{I}}$}}$ & M$_{\text{Fe$_{\text{II}}$}}$ & M$_{X^{\prime}}$ & M$_{Si}$  & P($\%$)  \\ [0.1ex]
\hline
Fe$_2$YSi & 	 23 & 	 0.96 & 	 0.54 & 	 0.53 & 	 -0.05 & 	 -0.02 & 	 89 \\ 
Fe$_2$ZrSi & 	 24 & 	 0.00 & 	 0.00 & 	 0.00 & 	 -0.00 & 	 -0.00 & 	 0 \\ 
Fe$_2$NbSi & 	 25 & 	 0.97 & 	 0.60 & 	 0.60 & 	 -0.14 & 	 -0.01 & 	 95 \\ 
Fe$_2$MoSi & 	 26 & 	 1.79 & 	 0.96 & 	 0.96 & 	 -0.05 & 	 -0.01 & 	 88 \\  [1ex]
\hline
Systems(T$_\text{II}$)& N$_{\text{V}}$ & M     & M$_{\text{Fe$_{\text{I}}$}}$ & M$_{X^{\prime}}$ & M$_{\text{Fe$_{\text{II}}$}}$ &   M$_{Si}$  & P($\%$)   \\ [0.1ex]
\hline
Fe$_2$TcSi & 	 27 & 	 3.78 & 	 1.37 & 	 -0.05 & 	 2.43 & 	 0.01 & 	 16 \\ 
Fe$_2$RuSi & 	 28 & 	 4.69 & 	 1.65 & 	 0.34 & 	 2.74 & 	 -0.02 & 	 8 \\ 
Fe$_2$RhSi & 	 29 & 	 4.90 & 	 1.72 & 	 0.38 & 	 2.85 & 	 -0.03 & 	 71 \\ 
Fe$_2$PdSi & 	 30 & 	 5.07 & 	 2.08 & 	 0.18 & 	 2.79 & 	 -0.02 & 	 31 \\ 
Fe$_2$AgSi & 	 31 & 	 4.69 & 	 2.07 & 	 0.00 & 	 2.62 & 	 -0.02 & 	 46 \\   [1ex]

\hline\hline
\end{tabular}
\end{table}
\begin{table}
\caption{\label{table7}Total and atomic magnetic moment of Co$_{2}$X$^{\prime}$Al systems in $\mu_{B}/f.u.$.  N$_{\text{V}}$ is the number of valence electron of the systems. M is the total moment and M$_{\text{i}}$ is the moment of constituent $\text{i}$.}
\vspace{2 mm}
\centering
\begin{tabular}{ c c c c c c c c c c}
\hline\hline
Systems(T$_\text{I}$) &  N$_{\text{V}}$ & M     & M$_{\text{Co$_{\text{I}}$}}$ & M$_{\text{Co$_{\text{II}}$}}$ & M$_{X^{\prime}}$ & M$_{Al}$  & P($\%$)  \\ [0.1ex]
\hline
Co$_2$YAl & 	 24 & 	 0.00 & 	 0.00 & 	 0.00 & 	 0.00 & 	 0.00 & 	 0 \\ 
Co$_2$ZrAl & 	 25 & 	 1.00 & 	 0.59 & 	 0.59 & 	 -0.10 & 	 -0.00 & 	 100 \\ 
Co$_2$NbAl & 	 26 & 	 2.00 & 	 1.02 & 	 1.02 & 	 0.01 & 	 0.00 & 	 100 \\ 
Co$_2$MoAl & 	 27 & 	 2.84 & 	 1.19 & 	 1.19 & 	 0.47 & 	 -0.01 & 	 4 \\ 
Co$_2$TcAl & 	 28 & 	 3.91 & 	 1.36 & 	 1.36 & 	 1.25 & 	 -0.03 & 	 56 \\ [1ex]
\hline
Systems(T$_\text{II}$)& N$_{\text{V}}$ & M     & M$_{\text{Co$_{\text{I}}$}}$ & M$_{X^{\prime}}$ & M$_{\text{Co$_{\text{II}}$}}$ &   M$_{Al}$  & P($\%$)   \\ [0.1ex]
\hline
Co$_2$RuAl & 	 29 & 	 3.87 & 	 1.32 & 	 0.70 & 	 1.94 & 	 -0.01 & 	 69 \\ 
Co$_2$RhAl & 	 30 & 	 3.37 & 	 1.31 & 	 0.39 & 	 1.80 & 	 -0.02 & 	 79 \\ 
Co$_2$PdAl & 	 31 & 	 3.04 & 	 1.46 & 	 0.04 & 	 1.63 & 	 -0.01 & 	 54 \\ 
Co$_2$AgAl & 	 32 & 	 2.77 & 	 1.34 & 	 -0.02 & 	 1.56 & 	 -0.02 & 	 64 \\  [1ex]

\hline\hline
\end{tabular}
\end{table}

\begin{table}
\caption{\label{table8}Total and atomic magnetic moment of Co$_{2}$X$^{\prime}$Al systems in $\mu_{B}/f.u.$.  N$_{\text{V}}$ is the number of valence electron of the systems. M is the total moment and M$_{\text{i}}$ is the moment of constituent $\text{i}$.}
\vspace{2 mm}
\centering
\begin{tabular}{ c c c c c c c c c c}
\hline\hline
Systems(T$_\text{I}$) &  N$_{\text{V}}$ & M     & M$_{\text{Co$_{\text{I}}$}}$ & M$_{\text{Co$_{\text{II}}$}}$ & M$_{X^{\prime}}$ & M$_{Si}$  & P($\%$)  \\ [0.1ex]
\hline
Co$_2$YSi & 	 25 & 	 1.00 & 	 0.58 & 	 0.58 & 	 -0.12 & 	 0.01 & 	 100 \\ 
Co$_2$ZrSi & 	 26 & 	 2.00 & 	 1.04 & 	 1.04 & 	 -0.07 & 	 0.04 & 	 100 \\ 
Co$_2$NbSi & 	 27 & 	 1.71 & 	 0.83 & 	 0.83 & 	 0.04 & 	 0.03 & 	 72 \\ [1ex]
\hline
Systems(T$_\text{II}$)& N$_{\text{V}}$ & M     & M$_{\text{Co$_{\text{I}}$}}$ & M$_{X^{\prime}}$ & M$_{\text{Co$_{\text{II}}$}}$ &   M$_{Si}$  & P($\%$)   \\ [0.1ex]
\hline
Co$_2$MoSi & 	 28 & 	 0.01 & 	 0.02 & 	 -0.00 & 	 -0.01 & 	 -0.00 & 	 0 \\ 
Co$_2$TcSi & 	 29 & 	 0.21 & 	 0.04 & 	 0.02 & 	 0.14 & 	 0.00 & 	 22 \\ 
Co$_2$RuSi & 	 30 & 	 3.20 & 	 1.12 & 	 0.48 & 	 1.67 & 	 -0.01 & 	 74 \\ 
Co$_2$RhSi & 	 31 & 	 3.27 & 	 1.34 & 	 0.44 & 	 1.56 & 	 -0.01 & 	 65 \\   [1ex]
\hline
Systems(T$_\text{I}$) &  N$_{\text{V}}$ & M     & M$_{\text{Co$_{\text{I}}$}}$ & M$_{\text{Co$_{\text{II}}$}}$ & M$_{X^{\prime}}$ & M$_{Si}$  & P($\%$)  \\ [0.1ex]
\hline
Co$_2$PdSi & 	 32 & 	 2.35 & 	 1.17 & 	 1.17 & 	 0.11 & 	 -0.04 & 	 69 \\ 
Co$_2$AgSi & 	 33 & 	 0.20 & 	 0.12 & 	 0.12 & 	 -0.01 & 	 -0.01 & 	 18 \\ [1ex]

\hline\hline
\end{tabular}
\end{table}

\subsection{Exchange interactions and Curie temperature}
\begin{figure}
\centerline{\hfill
\psfig{file=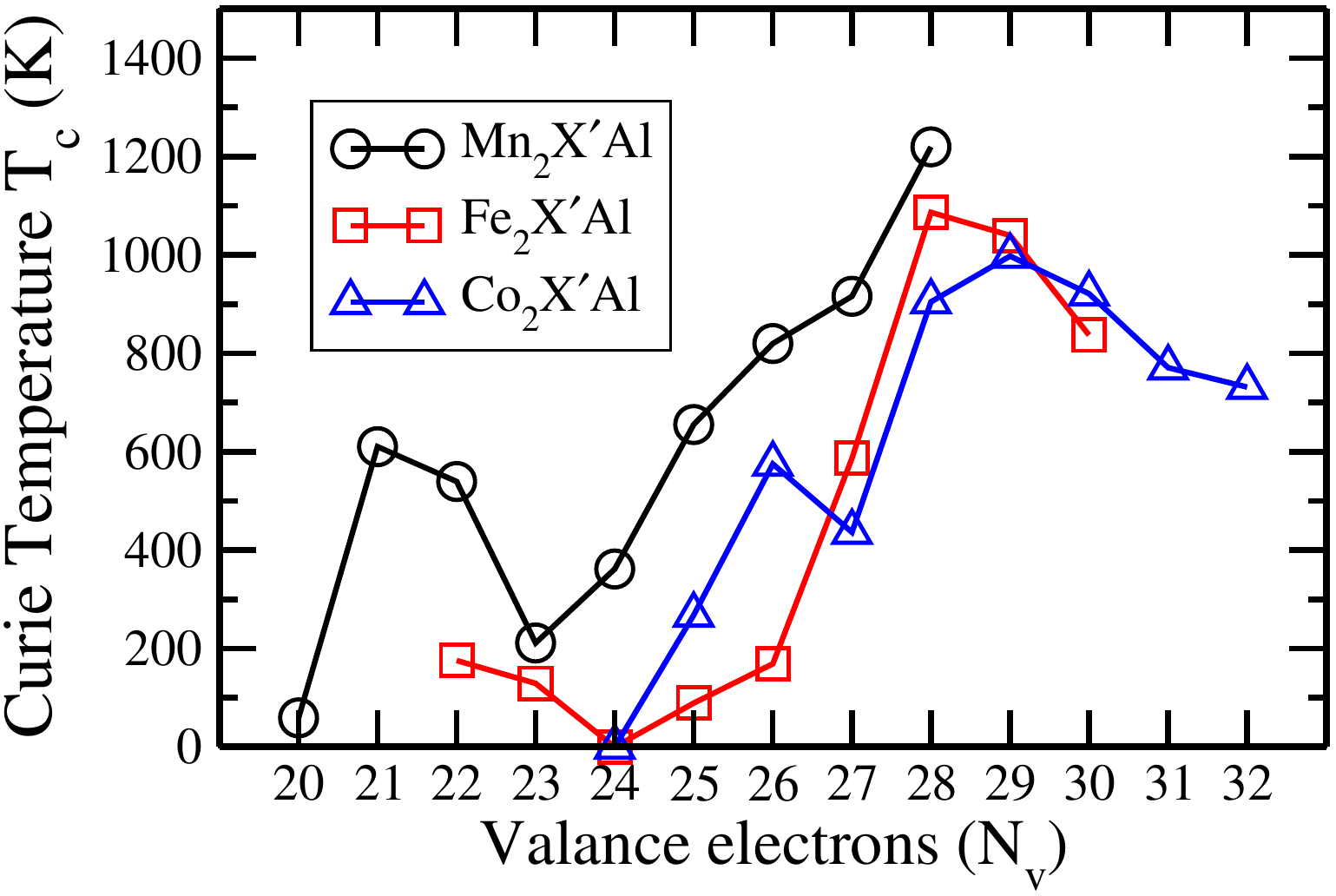,width=0.4\textwidth}\hfill}
 \caption{Variations in the Calculated Curie temperature with total number of valence electron for X$_2$X$^\prime$Al (X = Mn, Fe, Co) series.}
\label{curie-temp-x2yal}
\end{figure}

\begin{figure}
\centerline{\hfill
\psfig{file=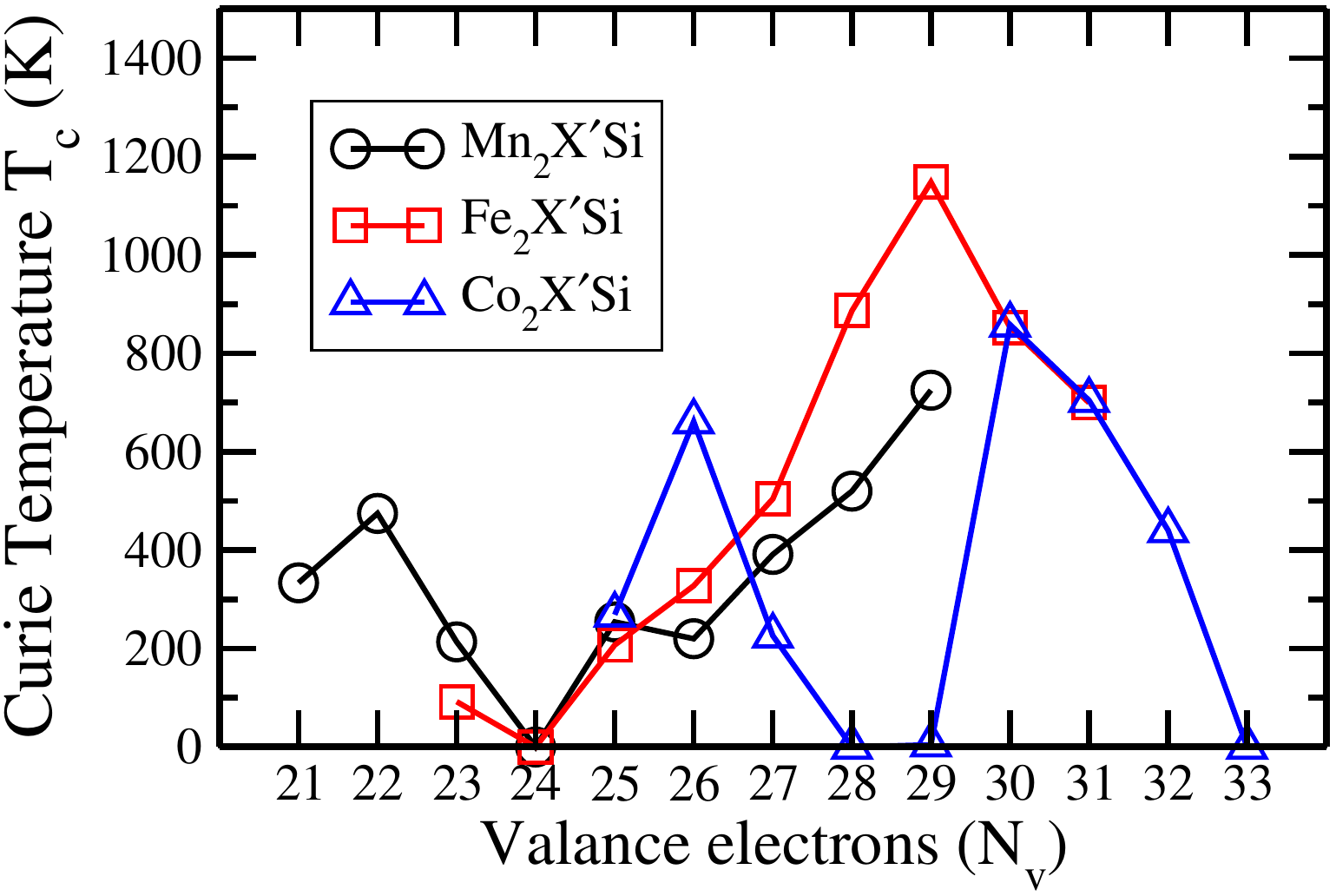,width=0.4\textwidth}\hfill}
 \caption{Variations in the calculated Curie temperature with total number of valence electron for  X$_2$X$^\prime$Si series.}
\label{curie-temp-x2ysi}
\end{figure}

In Fig \ref{curie-temp-x2yal} and \ref{curie-temp-x2ysi} we show the variations of Curie temperatures with  with changes in the valance electron number for X$_2$X$^\prime$Al (X = Mn, Fe, Co) and X$_2$X$^\prime$Si (X = Mn, Fe, Co) series respectively. Our results show that the variations in the Curie temperatures can be classified in two distinct regions based on structure types. Though the region-wise variations are not uniform it gives a qualitative idea of dependence of Curie temperatures on the ordering of the atoms.  we analyse  the trends in the variations of Curie temperatures by inspecting the variations of different effective exchange parameters $J^{eff}_{\mu\nu}$ ( Fig.11 and Fig.12 of supplementary material) being given as $J^{eff}_{\mu\nu}= \sum_{j} J^{0j}_{\mu\nu}$; $0$ fixed on $\mu$ sub-lattice and $j$ runs over $\nu$ sub-lattice. 

For Mn$_{2}$X$^{\prime}$Z compounds with structure type T$_{\text{I}}$, the Mn-Mn ferromagnetic interactions decide the variations in the Curie temperature T$_{\text{c}}$. This is true for Fe$_{2}$X$^{\prime}$Z and Co$_{2}$X$^{\prime}$Z series with compounds crystallising in T$_{\text{I}}$. The highest T$_{\text{c}}$ is thus obtained for Mn$_{2}$ZrAl and Mn$_{2}$ZrSi for Mn-series, Fe$_{2}$YAl and Fe$_{2}$MoSi for Fe-series, Co$_{2}$NbAl and Co$_{2}$ZrSi for Co-series. Among them, the Co-compounds have the highest T$_{\text{c}}$ and the Fe-compounds have the lowest T$_{\text{c}}$. This trend can be understood from the higher $J^{eff}$ for Co-compounds and relatively lower ones in Fe-compounds. In case of Co-compounds, along with Co-Co, Co-X$^{\prime}$ ferromagnetic interactions too play a significant role in deciding T$_{\text{c}}$. We find that both $J^{eff}$ are maximum for N$_{\text{V}}$=26 across the Z series for Co-compounds. This can be attributed to  the changes in the behaviour of the Exchange average, the average of exchange energies associated with low temperature spin excitations. This variation in the Exchange average is related to the availability of spin down states below Fermi level. A gap in the spin down bands starting below Fermi level and extending beyond would lead to a larger value of Exchange average and consequently a larger Curie temperature \cite{KublerPRB07}. This, exactly, is happening for Co-compounds considered here.
The same explanation can  be used for Mn$_{2}$- compounds. The absence of such clear gaps around Fermi level of Fe-compounds imply that this argument cannot be used and thus the trends in variations of T$_{\text{c}}$ is qualitatively different.

The dominant exchange interactions for compounds with structure type  T$_{\text{II}}$  are Mn$_{\text{I}}$(Co$_{\text{I}}$,Fe$_{\text{I}}$)-Mn$_{\text{II}}$(Co$_{\text{II}}$,Fe$_{\text{II}}$), Mn$_{\text{II}}$-Mn$_{\text{II}}$ (Co$_{\text{II}}$-Co$_{\text{II}}$, Fe$_{\text{II}}$-Fe$_{\text{II}}$)  and Mn$_{\text{II}}$-X$^{\prime}$(Co$_{\text{II}}$-X$^{\prime}$, Fe$_{\text{II}}$-X$^{\prime}$). For Mn-compounds, the Mn$_{\text{I}}$-Mn$_{\text{II}}$ interactions are strongly antiferromagnetic while the interactions between the 3$d$ magnetic atoms are ferromagnetic for compounds in other two series. In case of Mn-compounds, the T$_{\text{c}}$ for Si-series is generally less than Al-series. The reason is that in the compounds in Si series there are competing antiferromagnetic and ferromagnetic interactions with the ferromagnetic ones coming from Mn$_{\text{II}}$-Mn$_{\text{II}}$ and Mn$_{\text{II}}$-X$^{\prime}$ interactions. The variations in the T$_{\text{c}}$ for compounds in each series are controlled by the variations in the various dominant $J^{eff}$. 

\subsection{Trends in the electronic properties upon replacing a 3$d$ X$^{\prime}$ constituent with an isoelectronic 4$d$ element }
There is lot more half-metals and SGS discovered among X$_{2}$X$^{\prime}$Z Heusler compounds with all transition metal atoms from 3$d$ series. It would, thus, be interesting and important to understand the impact of replacing a 3$d$ X$^{\prime}$ constituent with an isoelectronic 4$d$ element. A comparison into the half-metallic aspects of the isoelectronic compounds which are either half-metals or near half-metals with high spin polarisation in either or both series (one with X$^{\prime}$ a 3$d$ element and another with X$^{\prime}$ an isoelectronic 4$d$ element) would shed light to the following: (a) whether electronic properties like half-metallicity primarily depends on N$_{\text{V}}$ and not on whether X$^{\prime}$ is a 3$d$ element or a 4$d$ element and (b) if not, then which are the possible factors that can explain the comparative trends in the isoelectronic compounds where in each case a 3$d$ X$^{\prime}$ is replaced with it's isoelectronic 4$d$ counterpart. Addressing this issue would help in prediction and design of new half-metals.
\begin{figure}[h!]
\centering
\subfigure[]{\includegraphics[scale=0.15]{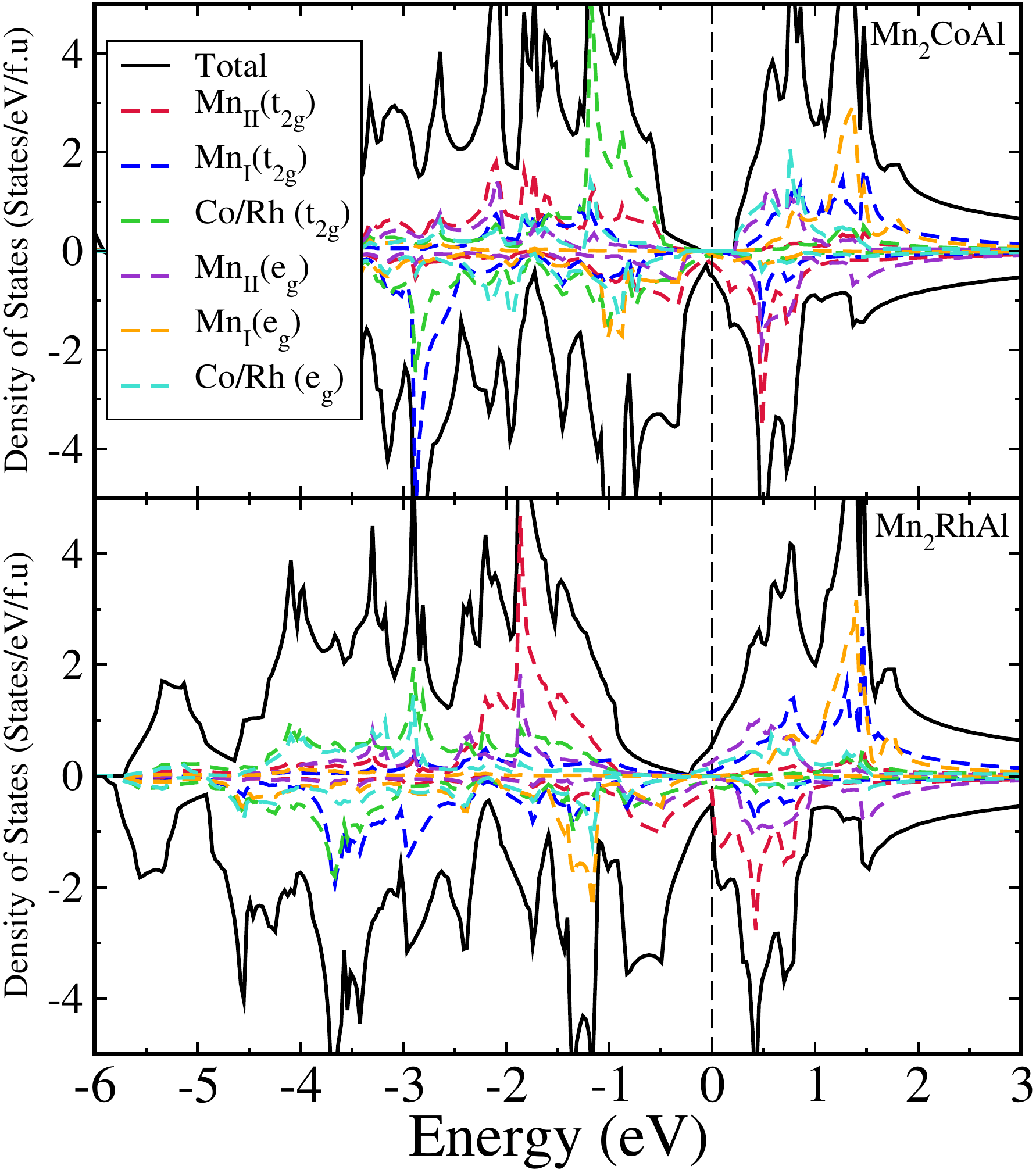}}
\subfigure[]{\includegraphics[scale=0.15]{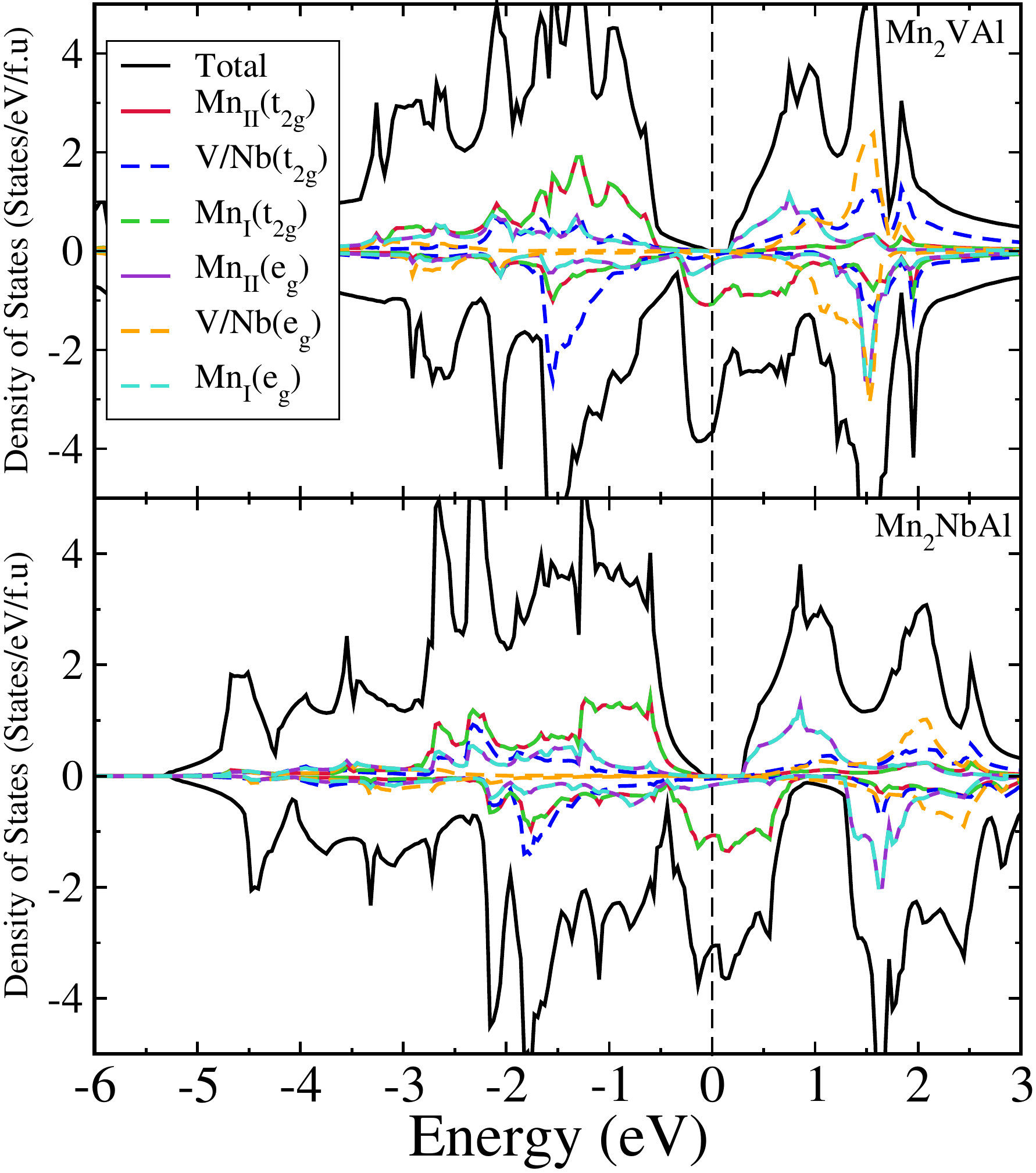}}
\subfigure[]{\includegraphics[scale=0.15]{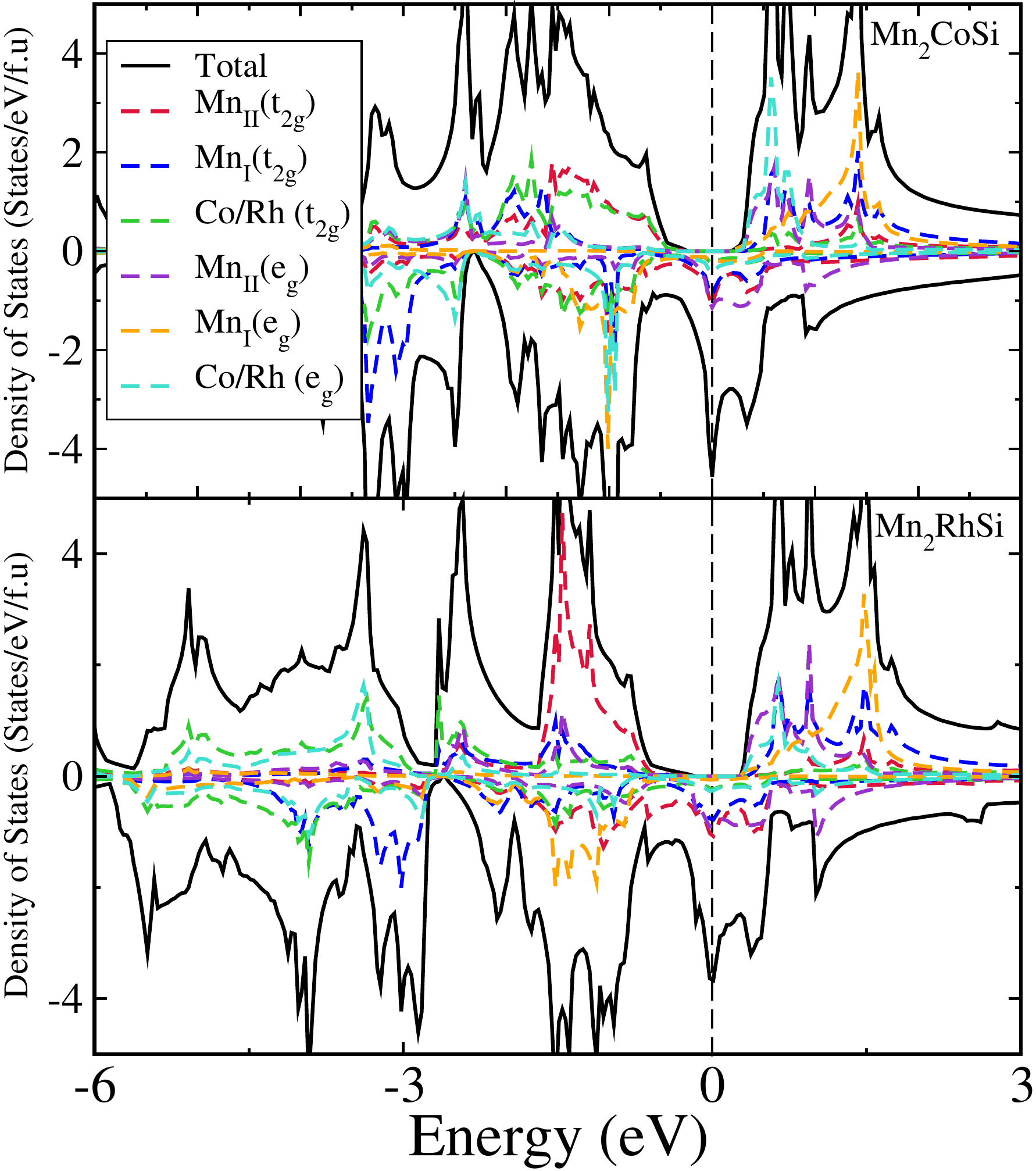}}
\subfigure[]{\includegraphics[scale=0.15]{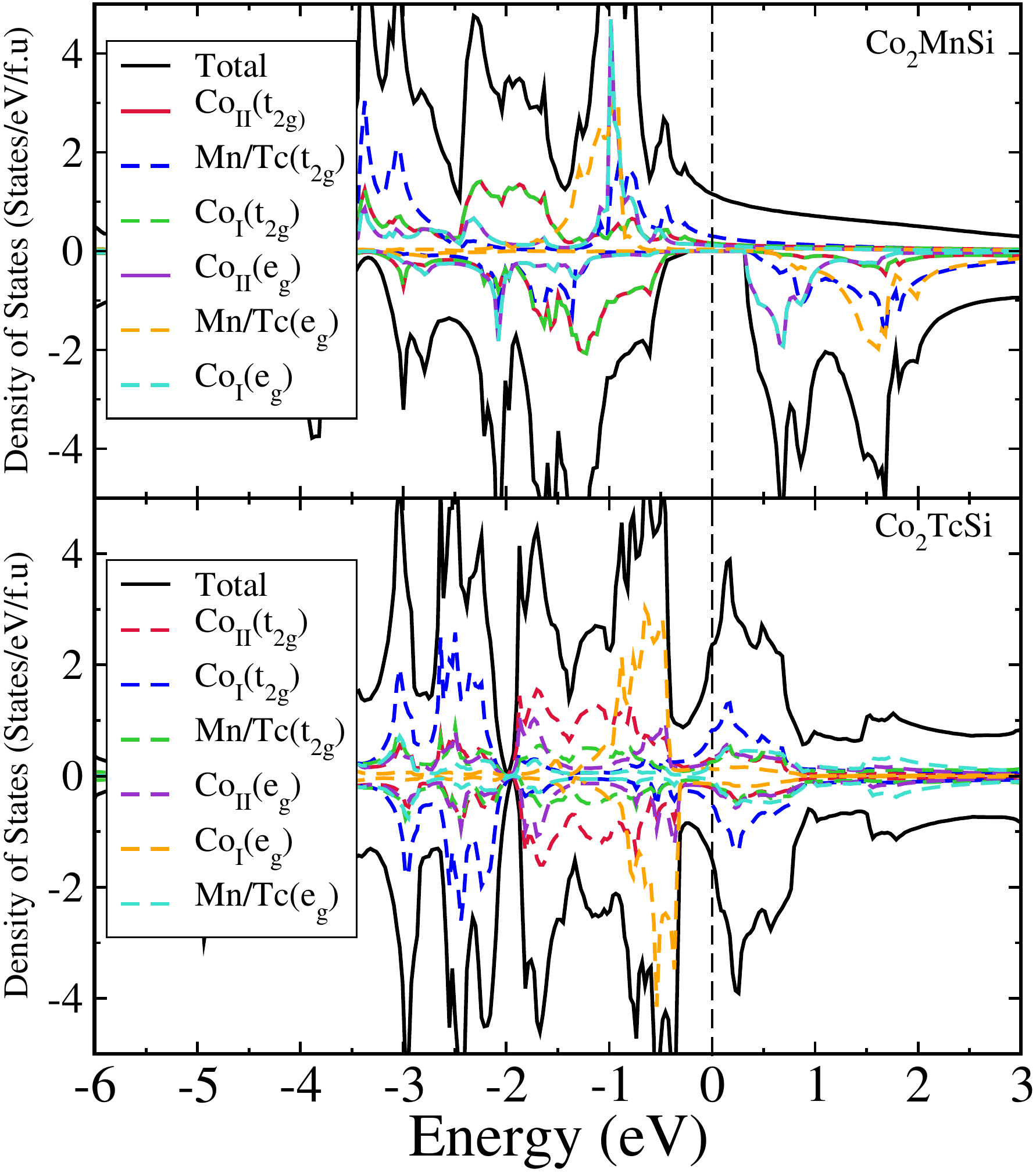}}
\subfigure[]{\includegraphics[scale=0.15]{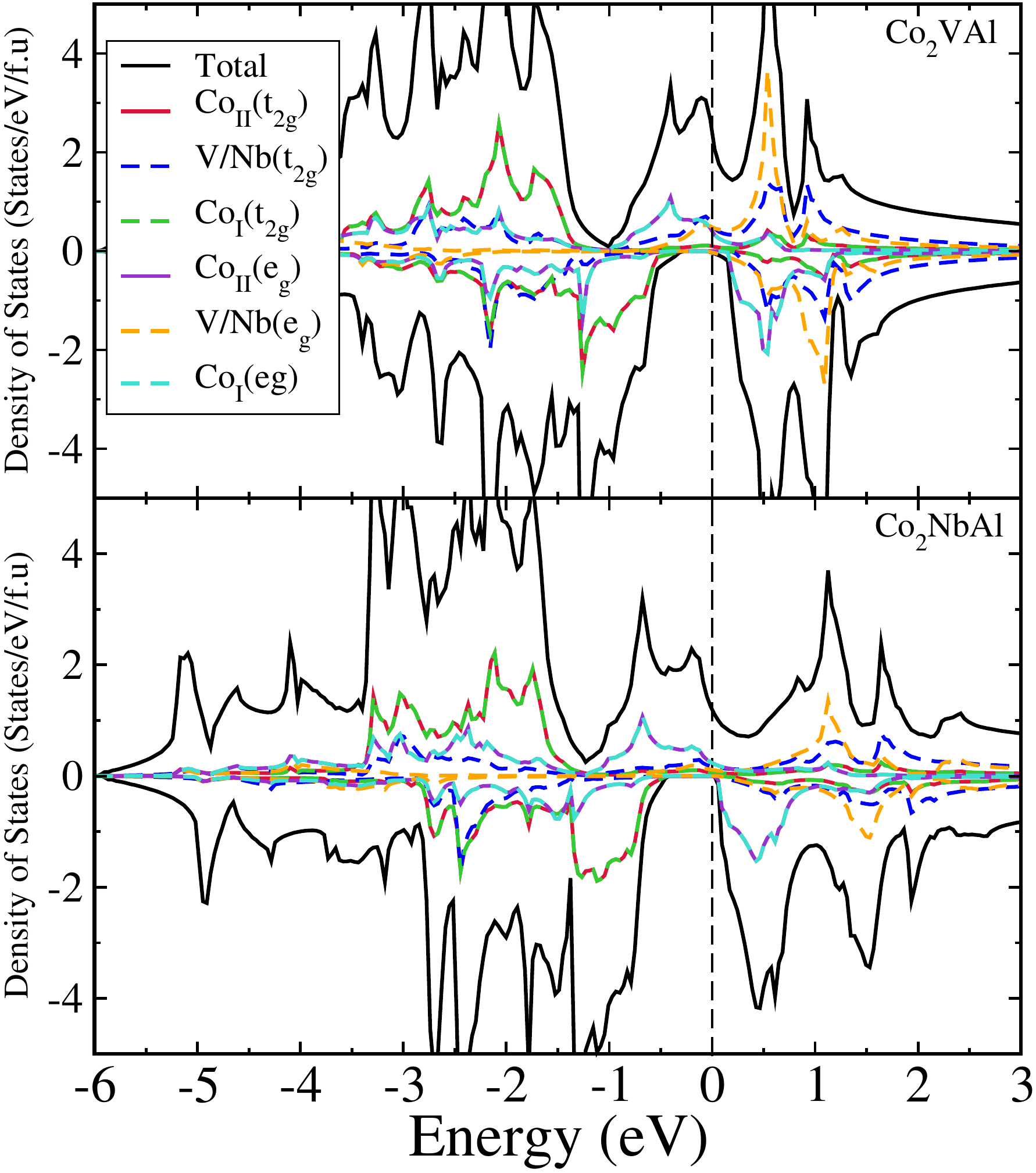}}
\subfigure[]{\includegraphics[scale=0.15]{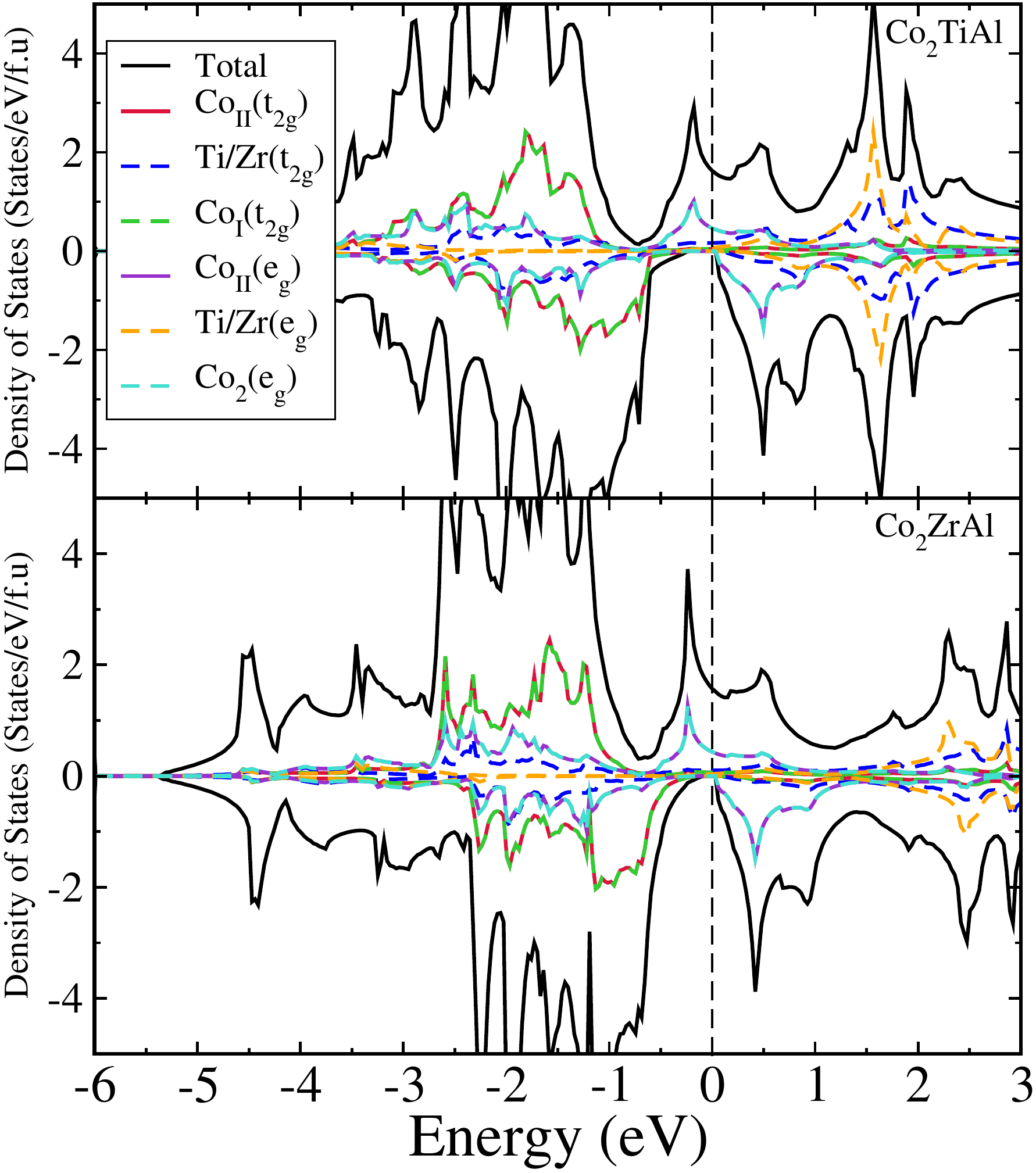}}
\caption{Spin polarized total and atom-projected densities of states for pairs of isoelectronic compounds({\bf a})Mn$_2$CoAl-Mn$_2$RuAl  ({\bf b}) Mn$_2$VAl-Mn$_2$NbAl ({\bf c})Mn$_2$CoSi-Mn$_2$RhSi ({\bf d}) Co$_2$MnSi-Co$_2$TcSi ({\bf e})Co$_2$VAl-Co$_2$NbAl and ({\bf f})Co$_2$TiAl-Co$_2$ZrAl.}
\label{DOSx2yZ}
\end{figure}

In Table I of supplementary material, the structural, magnetic and electronic properties associated with half-metallicity of X$_{2}$X$^{\prime}$Z (X=Co,Mn,Fe; X$^{\prime}$= an element with 3$d$ electrons; Z=Al,Si) obtained from various existing resources are tabulated. Focusing solely on the compounds with integer or near integer magnetic moments and spin polarisation equal to or close to 100$\%$ , we find that Mn$_{2}$VAl and Mn$_{2}$CoSi are half-metals, Mn$_{2}$TiSi and Mn$_{2}$FeAl are near half-metals while Mn$_{2}$CoAl is a SGS; Fe$_{2}$MnSi is a half-metal while Fe$_{2}$CrAl and Fe$_{2}$CrSi are near half-metals; Co$_{2}$TiAl,Co$_{2}$CrAl, Co$_{2}$ScSi and Co$_{2}$CrSi are half-metals while Co$_{2}$VAl is a near half-metal. For each of these compounds, we look at the spin polarisations of the isoelectronic counterpart in the series with X$^{\prime}$ a 4$d$ element and find the following: (i) Mn$_{2}$VAl-Mn$_{2}$NbAl and Mn$_{2}$CoSi-Mn$_{2}$RhSi pairs are half-metals, Mn$_{2}$FeAl-Mn$_{2}$RuAl pairs are near half-metals with high spin polarisation, Mn$_{2}$TiSi is a near half metal while isoelectronic Mn$_{2}$ZrSi is a half-metal, Mn$_{2}$CoAl in structure type T$_{\text{II}}$ is a SGS while Mn$_{2}$RhAl is an ordinary metal, (ii) Fe$_{2}$MoAl and Fe$_{2}$MoSi are ordinary metals while their isoelectronic counterparts with X$^{\prime}$ a 3$d$ element, Fe$_{2}$CrAl and Fe$_{2}$CrSi, respectively are compounds with high spin polarisations. Fe$_{2}$TcSi is an ordinary metal while Fe$_{2}$MnSi is a half-metal, (iii) Co$_{2}$TiAl-Co$_{2}$ZrAl and Co$_{2}$ScSi-Co$_{2}$YSi are half-metal pairs, Co$_{2}$CrAl and Co$_{2}$MnSi are half-metals while their isoelectronic counterparts Co$_{2}$MoAl and Co$_{2}$TcAl, respectively, are ordinary metals with low spin polarisations,Co$_{2}$VAl is a near half metal while Co$_{2}$NbAl is a half metal. Co$_{2}$TiSi is an ordinary metal while Co$_{2}$ZrSi is a half-metal. Co$_{2}$CrSi is a half-metal while Co$_{2}$MoSi is a non-magnetic material.

In order to understand the trend observed, we take recourse to the comparisons of the electronic structures of each pairs of compounds considered above. In Fig. \ref{DOSx2yZ} and in Figs. 13-17 of supplementary material we show the total and partial densities of states of some of the pairs. In cases of Mn$_{2}$CoSi-Mn$_{2}$RhSi and Mn$_{2}$VAl-Mn$_{2}$NbAl pairs (Fig. \ref{DOSx2yZ} (b),(c)), we find that the electronic structures of a given pair are near identical around the Fermi levels. In cases of inverse Heusler (structure type T$_{\text{II}}$) Mn$_{2}$CoSi and Mn$_{2}$RhSi, the half-metallic gaps are artefacts of the separation of the t$_{2g}$ and e$_{g}$ spin up states. However, there is a difference between the contributors to these states between the two compounds: In Mn$_{2}$RhSi, the states bordering the gap are primarily Mn$_{\text{I}}$ t$_{2g}$ and a hybridised Mn$_{\text{II}}$-Rh e$_{g}$ while in Mn$_{2}$CoSi, the t$_{2g}$ states too are hybridised Mn$_{\text{I}}$-Co. In case of the pair Mn$_{2}$VAl-Mn$_{2}$NbAl, both in regular Heusler structure (structure type T$_{\text{I}}$), the electronic structures in the majority band is identical-the half-metallic gap being flanked by Mn states only. In Fig. 13 of supplementary material, we show the densities of states of the pair Mn$_{2}$TiSi-Mn$_{2}$ZrSi, the later is a half-metal while the former is a near half-metal with integer moment and above 90$\%$ spin polarisation (Table I, supplementary material). The densities of states reveal that the difference in the electronic properties originates from the positions of the t$_{2g}$ and e$_{g}$ states flanking the half-metallic gap; in Mn$_{2}$TiSi, the position of the top of the t$_{2g}$ spin up bands in the occupied part coming from the Mn atoms cut through the Fermi level while this is not so in case of Mn$_{2}$ZrSi. The little trace of densities of states at the Fermi level reduces the spin polarisation in Mn$_{2}$TiSi. The striking difference in electronic properties in the context of half-metallicity is observed in the pair Mn$_{2}$CoAl-Mn$_{2}$RhAl with the former being a SGS and both are of structure type T$_{\text{II}}$. A close inspection into the densities of states (Fig. \ref{DOSx2yZ} (a)), however, reveals that the electronic structures of the two compounds are not so different from each other-in both compounds top of the valence band and bottom of the conduction band touch each other forming zero gap in the spin down band. A difference occurs in the spin up bands. In Mn$_{2}$CoAl, there is a clear half-metallic gap with the top of the valence band having contributions from Mn$_{\text{II}}$ and Co t$_{2g}$ states and the bottom of the conduction band having contributions from the e$_{g}$ states of the same pair of atoms. In Mn$_{2}$RhAl, we find another zero gap in the spin up bands, however, not at the fermi level but inside the occupied part of the spectrum. The t$_{2g}$ and e$_{g}$ bands in this case are more delocalised, possibly due to hybridisations with more delocalised 4$d$ states of Rh as compared to 3$d$ states of Co, thus, diminishing chances of getting SGS or half-metallic properties in this compound. 

The origin of different properties in Fe$_{2}$MnSi-Fe$_{2}$TcSi pair can be traced to the structures they crystallise in: the former in T$_{\text{I}}$ and the later in T$_{\text{II}}$. In Fig. 14 of supplementary material, we compare the densities of states of two compounds. The half-metallic behaviour of Fe$_{2}$MnSi stems from the unavailability of spin down states on either side of Fermi level with the gap being flanked by Fe $t$ and $e$ states. In Fe$_{2}$TcSi, the T$_{\text{II}}$ structure leaves no such scope as there is substantial hybridisations between Fe and Tc states across the Fermi level. The pairs Fe$_{2}$CrAl-Fe$_{2}$MoAl and Fe$_{2}$CrSi-Fe$_{2}$MoSi all crystallise in same structure (type T$_{\text{I}}$) but their spin polarisations differ significantly. However, the densities of states of each pair of compounds are qualitatively similar. The highlights of the electronic structure in both Fe$_{2}$CrAl \cite{Fe2CrAl_4} and Fe$_{2}$MoAl are a pseudogap in the spin up band and a gap cutting through the Fermi level in the spin down band, although in case of Fe$_{2}$MoAl, it is more like a pseudo gap reducing it's spin polarisation. This difference in the electronic structure once again stems from the availability of states near and at Fermi level in case of Fe$_{2}$MoAl, presumably because of the hybridisations of Fe with delocalised Mo states. Similar differences explain the differences in spin polarisations of Fe$_{2}$CrSi and Fe$_{2}$MoSi.

In Fig. \ref{DOSx2yZ} (d)-(f), we show the densities of states for three pairs of Co$_{2}$-compounds. The Co$_{2}$TiAl-Co$_{2}$ZrAl pair is a half-metal one, in Co$_{2}$VAl-Co$_{2}$NbAl pair, the later is a half-metal while the former is a near half-metal with spin polarisation of 95$\%$. In the pair Co$_{2}$MnSi-Co$_{2}$TcSi, the former is a half-metal and the later an ordinary metal. In Figs. 15-17 of supplementary material we show the electronic structures of another three pairs Co$_{2}$CrAl-Co$_{2}$MoAl, Co$_{2}$ScSi-Co$_{2}$YSi and Co$_{2}$TiSi-Co$_{2}$ZrSi respectively. We find that the electronic structures of Co$_{2}$TiAl and Co$_{2}$ZrAl are identical with the Co states on either sides of the gaps being located even at same energies. Same happens in case of Co$_{2}$ScSi-Co$_{2}$YSi pair. The differences in the electronic structures near the Fermi levels of Co$_{2}$VAl and Co$_{2}$NbAl are minimal. Unlike Co$_{2}$NbAl, there is a trace of hybridisation between Co and V states around the Fermi level reducing the spin polarisation by a few percent than that of an ideal half-metal. A major difference is observed between Co$_{2}$MnSi and Co$_{2}$TcSi. This, clearly, is due to the different crystal structures (Co$_{2}$MnSi has structure type T$_{\text{I}}$, Co$_{2}$TcSi has structure type T$_{\text{II}}$), the reflections are there in the densities of states. Similar is the case with Co$_{2}$CrSi \cite{Co2CrSi_1} and Co$_{2}$MoSi. Analysing the cases where both members of an isoelectronic pair crystallise in structure type T$_{\text{I}}$ but one of them is a half-metal while the other is a metal with medium spin polarisation (Figs 15, 17 of supplementary material), we find that the electronic structures are very similar for the compounds in a pair. For example, in both Co$_{2}$TiSi-Co$_{2}$ZrSi and Co$_{2}$CrAl-Co$_{2}$MoAl pairs, the noteworthy feature in the densities of states is the presence of a clear gap in the spin down bands. The differences in the spin polarisation arise due to different positions of the Fermi levels. In Co$_{2}$TiSi, the bottom of the conduction band which consists of e$_{g}$ states of Co falls behind the Fermi level while they are above the Fermi level in Co$_{2}$ZrSi. Similar is the case with the other pair of compounds.


\section{Conclusions}
Employing first-principles electronic structure calculations we have systematically studied the structural, electronic and magnetic properties of 54 X$_2$X$^\prime$Z ternary Heusler compounds where X= Mn, Fe, Co; Z = Al, Si and   X$^\prime$  represents 9 elements with $4d$ electrons in their valance shells. In pursuit of finding new half-metallic magnets from ternary series with both $3d$ and $4d$ electrons in the same compounds we found only seven half-metals Mn$_2$NbAl, Mn$_2$ZrSi, Mn$_2$RhSi, Co$_2$ZrAl, Co$_2$NbAl, Co$_2$YSi and  Co$_2$ZrSi with 100\% spin polarisation. We also find the compounds Mn$_2$TcAl, Mn$_2$RuAl, Mn$_2$NbSi, Mn$_2$RuSi, Fe$_2$NbSi with high spin polarisation ( greater than 90\%) and a gap like feature in one of the spin-channel near the Fermi-level. These compounds can be identified as ``near half-metals". Tuning the positions of their Fermi levels by application of pressure such that the gap cuts through the Fermi levels may induce half-metallicity in them.

From the present study the hybridisation picture that has been in use to explain the origin of half-metallicity in ternary Heusler compounds with magnetic components being the ones with 3$d$ elements only, is found to be valid in case of compounds with both 3$d$ and 4$d$ elements as constituents. This leads to a general framework in understanding the origin of half-metallic behaviour in Heusler X$_{2}$X$^{\prime}$Z compounds where X$^{\prime}$ can be either a 3$d$ element or one from the group of 4$d$. An one-on-one comparison between X$_{2}$X$^{\prime}$Z compounds where X$^{\prime}$ is a 3$d$ element in one case and a 4$d$ element in other, shows that as long as N$_{\text{V}}$, the number of valence electrons are same, the electronic properties like half-metallicity or near half-metallicity with high spin polarisation, by an large, remain intact if a 4$d$ element replaces a 3$d$ one as X$^{\prime}$ provided both compounds crystallise in the same ground state structure. Even in cases where the spin polarisations, and subsequent half-metallic features are different, the electronic structures are very similar for compounds with same N$_{\text{V}}$ and same structure type. The differences in electronic properties in regard to half-metallicity occurs due to the relative positions of the bottom of the conduction band and the Fermi energy which is a consequence of the positions of the X and X$^{\prime}$ states near the Fermi level. Therefore, N$_{\text{V}}$ can be a good predictor to explore possible new half-metals.

The trends in the Curie temperature, T$_{\text{c}}$, in the compounds with X$^{\prime}$ a 4$d$ element, is found to be correlated with the trends in the dominant exchange interactions. We find that across the series, the X-X exchange interactions determine the trends in T$_{\text{c}}$ for compounds with structure type T$_{\text{I}}$ which in turn can be correlated to the dominance of X-X hybridisations in their electronic structures. In case of compounds with structure type T$_{\text{II}}$, the dominant exchange interactions are X$_{\text{I}}$-X$_{\text{II}}$ and X$_{\text{II}}$-X$^{\prime}$ deciding the trends in T$_{\text{c}}$. This, once again, can be correlated to the dominant hybridisations in their electronic structures. The present work, thus, systematically explores the physics behind occurrence of half-metallicity in compounds with magnetic constituents from both 3$d$ and 4$d$ series and provides with a generalised microscopic picture applicable to a large number of heusler compounds. This would be useful for experimentalists in particular, to explore new materials with novel magnetic applications.
%
\section*{ACKNOWLEDGMENT}

We thank Dr. Ashis Kundu for useful discussion and fruitful inputs through out this work. IIT Guwahati and DST India is acknowledged for providing the PARAM superconducting facility and the computer cluster in the Department of Physics, IIT Guwhati.

\bibliographystyle{aip} 
\bibliography{ref}

\end{document}